\documentclass[journal, letterpaper, twocolumn]{IEEEtran}
\usepackage{citesort}
\usepackage{subcaption}
\usepackage{kantlipsum}
\usepackage{enumitem}
\usepackage{supertabular}
\usepackage{longtable}
\usepackage{makecell}
\usepackage[hyphens]{url} 
\usepackage{rotating}
\usepackage{multirow}
\usepackage{gensymb}
\usepackage{array,ragged2e}
\usepackage{cite}
\usepackage{epsfig}
\usepackage{color}
\usepackage[latin9]{inputenc}
\usepackage{algorithmic}
\usepackage{algorithm}
\usepackage{mathrsfs}
\usepackage{mathtools}
\usepackage{tikz}
\usetikzlibrary{arrows,shadows}
\usepackage{pgf-umlsd}
\usepackage{caption}
\usepackage{comment}
\usepackage{xcolor}
\usepackage{amsmath,amssymb,amsfonts}
\usepackage{graphicx}
\usepackage{graphics}
\usepackage{textcomp}
\usepackage{mathtools}
\usepackage{array}
\usepackage{longtable}
\usepackage{stackengine}
\usepackage{xcolor}
\usepackage{cite}
\usepackage{float}

\usepackage[acronym,nonumberlist,toc]{glossaries}
\usepackage{glossary-superragged}

\newglossarystyle{modsuper}{%
  \setglossarystyle{super}%
  \renewcommand*{\printglossary}[5]{%
    \glsentryitem{##1}\glstarget{##1}{##2} & ##3\glspostdescription\space ##5\\[3pt]}%
}

\makeglossaries

\renewcommand*{\glspostdescription}{} 

\newacronym{snspd}{SNSPD}{Superconducting Nanowire Single Photon Detector}
\newacronym{gmapd}{GmAPD}{Geiger-Mode Avalanche Photodiode}
\newacronym{ingaas}{InGaAs}{Indium Gallium Arsenide}
\newacronym{pdu}{PDU}{Protocol Data Units}
\newacronym{qam}{QAM}{Quadrature Amplitude Modulation}
\newacronym{ofdm}{OFDM}{Orthogonal Frequency Division Multiplexing}
\newacronym{d2nn}{D2NN}{Diffractive Deep Neural
Network}
\newacronym{np}{NP}{Network Protocol}
\newacronym{dsql}{DSQL}{Deep Space Quantum Link}
\newacronym{au}{AU}{Astronomical Unit }
\newacronym{edrs}{EDRS}{European Data Relay System}
\newacronym{sps}{SPC}{Single Photon Coherent}
\newacronym{sq}{SQ}{Single Quadrature}
\newacronym{slm}{SLM}{Spatial Light Modulators}
\newacronym{bp}{BP}{Bundle Protocol}
\newacronym{cgr}{CGR}{Contact Graph Routing}
\newacronym{swap}{SWaP}{Size Weight and Power}
\newacronym{cgh}{CGH}{Computer Generated Hologram}
\newacronym{ds}{DS}{Deep Space}
\newacronym{ns}{NS}{Number State}
\newacronym{rf}{RF}{Radio Frequency}
\newacronym{fso}{FSO}{Free Space Optical communications}
\newacronym{psa}{PSA}{Phase Sensitive Optical Amplifiers}
\newacronym{bpqm}{BPQM}{Belief Propagation
with Quantum Messages}
\newacronym{dovg}{DOVG}{Dammann Optical Vortex Grating}
\newacronym{sdt}{SDT}{Superdense Teleportation Protocol}
\newacronym{pin}{PIN}{P-type Intrinsic N-type}

\newacronym{mm}{MM}{Mode Modulation}
\newacronym{pc}{PC}{Photon-Counting}
\newacronym{dsoc}{DSOC}{Deep Space Optical communications}
\newacronym{smm}{SMM}{spatial mode multiplexing}
\newacronym{nasa}{NASA}{National Aeronautics and Space Administration}
\newacronym{dsc}{DSC}{Deep Space communications}
\newacronym{mlcd}{MLCD}{Mars Laser communications Demonstration}
\newacronym{llcd}{LLCD}{Lunar Laser communications Demonstration}
\newacronym{dsn}{DSN}{Deep Space Network}
\newacronym{jpl}{JPL}{Jet Propulsion Laboratory}
\newacronym{dsif}{DSIF}{Deep Space Instrumentation Facility}
\newacronym{ppm}{PPM}{Pulse Position Modulation}
\newacronym{bpsk}{BPSK}{Binary Phase Shift Keying}
\newacronym{qpsk}{QPSK}{Quadrature Phase Shift Keying}
\newacronym{fsk}{FSK}{Frequency Shift Keying}
\newacronym{maven}{MAVEN}{Mars Atmospheric and Volatile EvolutioN}
\newacronym{psk}{PSK}{Phase Shift Keying}
\newacronym{sn}{SN}{Space Network}
\newacronym{maser}{MASER}{microwave amplification by stimulated emission of radiation}
\newacronym{wdma}{WDMA}{Wavelength Division Multiple Access}
\newacronym{snr}{SNR}{Signal to Noise Ratio}
\newacronym{qc}{QC}{Quantum Communications}
\newacronym{ber}{BER}{Bit Error Rate}
\newacronym{tcp}{TCP}{Transmission Control Protocol}
\newacronym{pnr}{PNR}{Photon Number Resolved}
\newacronym{ipn}{IPN}{Interplanetary Network}
\newacronym{ccsds}{CCSDS}{Consultative Committee for Space Data Systems}
\newacronym{iroc}{iROC}{Integrated Radio and Optical Communications}
\newacronym{cnn}{CNN}{Convolutional Neural Network}
\newacronym{rs}{RS}{Reed-Solomon}
\newacronym{nemo}{NEMO}{Network Mobility}
\newacronym{bsp}{BSP}{Basic Support Protocol}
\newacronym{scps}{SCPS}{Space Communications Protocol Specifications}
\newacronym{BSS}{bss}{Bundle Streaming Service}
\newacronym{spps}{SPP}{Spiral Phase Plate}
\newacronym{ip}{IP}{Internet Protocol}
\newacronym{dtn}{DTN}{Delay/Disruption Tolerant Networking}
\newacronym{pie}{PIE}{Photon Information Efficiency}
\newacronym{qpg}{QPG}{Quantum Pulse Gating}
\newacronym{dstp}{DS-TP}{Deep Space Transport Protocol}
\newacronym{ecc}{ECC}{Error Correction Code}
\newacronym{fom}{FoM}{Figure of Merit}
\newacronym{tp}{TP}{Transport Protocol}
\newacronym{arq}{ARQ}{Automatic Retransmission ReQuest}
\newacronym{dttp}{DTTP}{Delay Tolerant Transport Protocol}
\newacronym{dtpc}{DTPC}{Delay Tolerant Payload Conditioning}
\newacronym{hpe}{HPE}{High Photon Efficiency}
\newacronym{scppm}{SCPPM}{Serially Concatenated Pulse Position Modulation}
\newacronym{lg}{LG}{Laguerre Gaussian}
\newacronym{sam}{SAM}{Spin Angular Momentum}
\newacronym{oam}{OAM}{Orbital Angular Momentum}
\newacronym{edfa}{EDFA}{Erbium-Doped Fibre Amplifier}
\newacronym{ppb}{PPB}{Photon Per Bit}
\newacronym{tdrss}{TDRSS}{Tracking and Data Relay Satellite System}
\newacronym{qkd}{QKD}{Quantum Key Distribution}
\newacronym{qt}{QT}{Quantum Teleportation}
\newacronym{qss}{QSS}{Quantum Secret Sharing}
\newacronym{fer}{FER}{Frame Error Rate}
\newacronym{qsdc}{QSDC}{Quantum Secure Direct communications}
\newacronym{ldpc}{LDPC}{Low Density Parity Check}
\newacronym{ook}{OOK}{On-Off Keying}
\newacronym{apd}{APD}{Avalanche Photo Diode}
\newacronym{cfdp}{CFDP}{File Delivery Protocol}
\newacronym{sctp}{SCTP}{Stream Control Transmission Protocol}
\newacronym{sinemo}{SINEMO}{Seamless IP Diversity Based Network Mobility }
\newacronym{ams}{AMS}{Asynchronous Message Service}
\newacronym{ltp}{LTP}{Licklider Transmission Protocol}
\newacronym{ipoc}{IPoC}{IP Over CCSDS Space Links}
\newacronym{sdlp}{SDLP}{Space Data Link Protocols}
\newacronym{ep}{EP}{Encapsulation Packets}
\newacronym{uslp}{USLP}{Unified Space Data Link Protocol}
\newacronym{aos}{AOS}{Advanced Orbiting Systems}
\newacronym{pro1}{Prox-1}{Proximity-1}
\newacronym{udp}{UDP}{User Datagram Protocol}
\newacronym{cla}{CLA}{Convergence Layer Adapters}
\newacronym{eid}{EID}{Endpoint Identifiers}
\newacronym{rsppm}{RSPPM}{Reed-Solomon Pulse Position Modulation}
\newacronym{uri}{URI}{Universal Resource Identifiers}
\newacronym{sp}{SP}{Space Packets}
\newacronym{eaodr}{EAODR}{Earliest Arrival Optimal Delivery Ratio}
\newacronym{nen}{NEN}{Near-Earth Network}
\newacronym{mro}{MRO}{Mars Reconnaissance Orbiter}
\newacronym{tgo}{TGO}{Trace Gas Orbiter}
\newacronym{marco}{MarCO}{Mars Cube One}
\newacronym{spp}{SPP}{Spiral Phase Plate}
\newacronym{at}{AT}{Atmospheric Turbulence}
\newacronym{tm}{TM}{Telemetry}
\newacronym{tc}{TC}{Telecommand}
\newacronym{mimo}{MIMO}{Multiple Input Multiple Output}
\newacronym{spar}{SPAR}{Synthetic Partial Aperture Receiving}
\newacronym{ccd}{CCD}{Charge Coupled Device}

\newcolumntype{P}[1]{>{\RaggedRight\arraybackslash}p{#1}}

\newcommand\Ou{%
  \mathrel{{\ooalign{\hss\raisebox{-0.5ex}{$-$}\hss\cr\raisebox{0.5ex}{$+$}}}}}
\def\delequal{\mathrel{\ensurestackMath{\stackon[1pt]{=}{\scriptstyle\Delta}}}}

\begin{document}

 \title{DS communications: Trends, Challenges, and Solutions} \title{Demystifying DS communications: State of the Art, Challenges, and Future Directions} 
 \title{ Demystifying DS communications: Advances,  Challenges, and Opportunities } 
 \title{Connecting the Universe: Enablers,  Challenges, and Mitigation Techniques}  
 

  \title{Deep Space Optical Communications: A Survey on Enabling Technologies, Challenges, and Mitigation Techniques}  

 \title{How Can Optical Communications Shape the Future of Deep Space Communications? A Survey}

\author{Sarah~Karmous,
        Nadia~Adem, Mohammed~Atiquzzaman, and Sumudu Samarakoon

\thanks{Sarah Karmous and Nadia Adem are with the Department of Electrical and Electronic Engineering University of Tripoli, Libya (e-mail: s.karmous,n.adem@uot.edu.ly). Nadia Adem is also with the Libyan Authority for Scientific Research, Tripoli, Libya and the Faculty of Information Technology and Electrical Engineering, University of Oulu, Finland. Mohammed Atiquzzaman is with the School of Computer Science, University of Oklahoma, USA (e-mail: atiq@ou.edu). Sumudu Samarakoon is with the Faculty of Information Technology and Electrical Engineering, University of Oulu, Finland (e-mail: Sumudu.Samarakoon@oulu.fi).}}

\maketitle

\begin{abstract}
With a large number of deep space (DS) missions anticipated by the end of this decade,  reliable and high-capacity DS communications  are needed more than ever. Nevertheless, existing technologies are far from meeting such a goal.
Improving current systems does not only require  engineering leadership, but also, very crucially, investigating  potential  
 technologies that overcome the unique challenges of ultra-long DS 
 links. 
To the best of our knowledge, there has not been any  comprehensive surveys of DS communications  technologies over the last decade. Free space optical  (FSO) is an emerging DS technology, proven to acquire lower communications systems  {size weight and power (SWaP)} and achieve a very high capacity compared to its counterpart radio frequency (RF), the currently used DS technology.    
In this survey, we discuss the pros and cons of deep space optical communications (DSOC) and review their physical and networking characteristics. Furthermore, we provide{, for the  first time,  thoughtful discussions} about {implementing} orbital angular momentum (OAM) and  quantum communications (QC) for DS.
{We elaborate on how these technologies among  other field advances including interplanetary network (IPN) and RF/FSO systems improve   reliability, capacity, and security.} 
This paper provides  a holistic survey {of} DSOC technologies gathering  247 fragmented pieces of literature and including novel  perspectives  aiming to set the stage for more developments in the field.  

\end{abstract}


\section{Introduction}

Space exploration offers an opportunity to understand the universe, resulting in endless benefits for human life {in} diverse aspects  such as technologies, transportation, environment, etc.~\cite{ISCEG2013}. There is  an increased investment from several worldwide governmental space agencies~\cite{NASA2021plan} as well as private companies~\cite{SpaceXMoon}  aiming to expand human presence in the Solar system~\cite{edwards2017update,ISECG2018roadmap}. Several missions have been recently conducted, e.g.~\cite{CNSA2021Tianwen,AmalTechSpec2021,ProvencesMars2020}, to  understand Mars's atmosphere and  search for life there. 
Many other missions are scheduled for the next several years, such as the Rosalind Franklin explorer program~\cite{Exomarsprogram2022}. 
There are also plans to send crews to Mars as early as 2029~\cite{FirsthumanlandingMars2026}. 
  
The success of these missions relies heavily on advances in different fields such as physics, astronomy, and telecommunications. 
Therefore, advances in communications technology are natural drivers for achieving these goals. 
\subsection{Motivation} 
\begin{figure}[!t]
\centering
\includegraphics[width=0.8\linewidth]{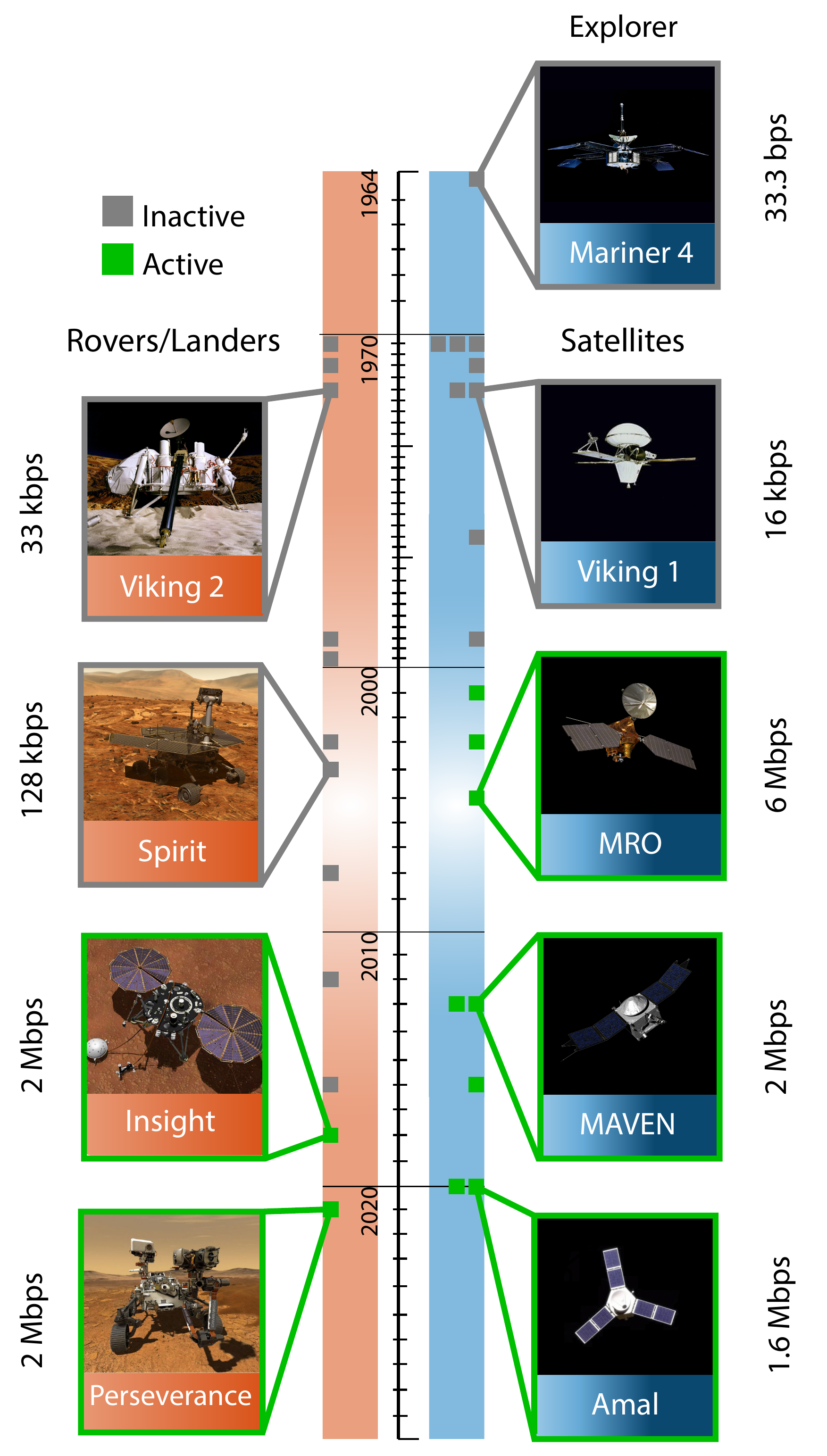}
\caption{Mars missions timeline with the data rates of their links to Earth.} 
\label{Fig.Timeline}
\end{figure}

Space exploration is built upon reliable broadband interplanetary communications~\cite{ESA20Mars}, which are only possible through developing dedicated technologies suitable for facing the unique challenges of deep space (DS) communications.
Unlike Earth and near-Earth, \acrshort{ds} communications suffer from  non-guaranteed line-of-sight, ultra-long time-changing distances, variability of Sun-Earth-probe angles, and {the coronal or Solar wind plasma,  referred to as Solar scintillation~\cite{xu2019effects}}. 
These challenges tremendously deteriorate the quality of the  links and hence, limit communications capacity~\cite{seas2019optical}. 
Current missions to Mars, for example, have only a few megabits per second communications rates at the minimum distance~\cite{arapoglou2017benchmarking,kwok2009DSNHandbook,edwards2018assessment}.  
Fig.~\ref{Fig.Timeline} shows a timeline for several Mars orbiters, landers, and rovers along with their data rates~\cite{AmalTechSpec2021,Russianspaceweb,MissiontoMars,ListofartificialMars,ListofMarsorbiters,JPL2021Mars,maevan101,NasaViking2,taylor2016mars,David2020Marina4,NasaViking1Orbiter,ProvencesMars2020}. 
It should be noted that higher attenuation is seen beyond Mars missions and the capacity drops by orders of magnitudes as the distance increases~\cite{AmalTechSpec2021}. 
For example, the Voyager 1 spacecraft,  which is located approximately {24.34 billion km from Earth  as of May 2024,}  operates  at a maximum data rate of a few kilo bits per second for less than a minute, only a few times a week~\cite{JplVoyage}. 

To build high-capacity DS communications, research efforts have been recently devoted to exploring the potentials of free space optical (FSO) communications~\cite{lyras2019deep,seas2019optical,hemmati2011deep,cesarone2011deep,kaushal2016optical,TrichiliJOSAB20}. Compared to radio frequency (RF), \acrshort{fso}  links enjoy larger bandwidth, smaller apertures at final stations, much improved security, less transmit power, and higher immunity against interference. {Although}  most recent  DS Mars missions are RF-based~\cite{edwards2018assessment}, 
  many projects have already adopted FSO 
  for Moon and Mars communications with plans to expand to other planets such as Jupiter~\cite{wilson1997results,biswas2006mars,grein2015optical,biswas2018deep,oh2019development}. %
 In addition,  several implementation-based studies have been done to evaluate the technology transition, e.g.~\cite{cesarone2011deep,deutsch2019creating}. Tremendous work is yet to be done though to make  FSO technology  
ready for future missions~\cite{fielhauer2012concurrent}.

Adopting FSO in DS  will play an important role in implementing interplanetary broadband networks.  
However, achieving such a goal will depend on the reliability, effectiveness, and efficiency of many other technologies, including modulation, coding, detection, and networking protocols.  
The criteria for selecting such schemes differ tremendously from those for terrestrial, airborne, and even near-Earth communications. 
In this survey, we highlight the various DS optical communications (DSOC)   design criteria and present the latest research and innovations in the area.

\subsection{Related Surveys}

There are few surveys available on  DS communications~\cite{hemmati2011deep,kaushal2016optical,biswas2018deep,wu2018overview,cesarone2011deep,de2011reliability,zhang2011survey,hall2020survey,konsgen2021current}, {most of which are  outdated and/or do not address the field comprehensively.}  
{Although}~\cite{hemmati2011deep} covers various aspects of DSOC,  an updated survey that reflects the current field demands, challenges, and  advances  is needed.
The survey of \cite{kaushal2016optical} covers some aspects of optical communications in DS, but  only partially.
Biswas et al.~\cite{biswas2018deep} {focus} merely on the progress made and the status of \acrshort{dsoc}  in National Aeronautics and Space Administration (\acrshort{nasa}) projects. 
In the same direction,~\cite{wu2018overview} overviews mainly laser-based DS demonstration projects and future missions.
Meanwhile, \cite{cesarone2011deep}  provides only  studies about  DS systems architects and discussions about the transition from  RF  to optical technology.
In addition,~\cite{de2011reliability} focuses only on DS networking aspects. 
Similarly,~\cite{zhang2011survey} covers merely several  aspects of RF-based DS communications systems. 
Furthermore, \cite{hall2020survey} addresses FSO and RF  challenges  and  discusses  DS systems implementations and demonstrations; however, it does not present research discussions and future perspectives  in the field.  
Finally,~\cite{konsgen2021current} gives an inspiring and relatively recent overview 
 of  DS networking and implementations,  yet, it  lacks any further details  in the other aspects, including physical characteristics and  field advances.  
 
{Despite their valuable contributions, none of the above works  covers DS communications   holistically.} 
In addition, most of these surveys are not recent. 
In contrast, as we demonstrate in Table \ref{Table:RelatedSurvComp}, our work aims to cover the key aspects of DSOC, add the missing areas from the aforementioned related work, and discuss the latest advances in the  field,  {including orbital angular momentum (\acrshort{oam}), interplanetary network (IPN), quantum communications (QC), and  the RF/FSO  which is an integration of both  RF and FSO technologies}.


\subsection{Contributions}
In this survey, we cover  existing DSOC systems' limitations,  recommended technologies to cope with them, and challenges facing their adoption. 
The contributions of this paper relative to the literature are summarized as follows.

\begin{enumerate}[label=\arabic*)]

\item We cover the major challenges of DS communications. We also discuss DSOC promises, challenges, and potential solutions and compare them with the  existing RF technology.


\item We survey the widely used modulation and coding schemes and the opportunities for enhancing performance by fusing them with  emerging detection techniques. 
 
 \item {We review recommended  networking architecture models, present  key characteristics of  promising protocols, and identify related open research directions.} 

\item  We present an overview of the IPN and study the visions for its future. 

\item  To the best of our knowledge, we provide the first thoughtful discussion about implementing OAM for DS and identify  challenges and potential solutions. 

\item We discuss the advantages, efforts, and challenges {of} 
 implementing~\acrshort{qc} and 
partnering RF and FSO technologies in DS. 
 
\end{enumerate}

With these contributions, we aim to pave the way for empowering more rapid  advancements toward   
  robust communications services in DS.

\begin{table*}[ht]
\scriptsize\centering

\caption{{COMPARISON OF DS COMMUNICATIONS SURVEYS}}
\label{Table:RelatedSurvComp}
\begin{tabular}{|c|c|c|c|c|c|c|c|c|c|c|c|}
\hline
Main theme              & Key aspects  & ~\cite{hemmati2011deep} &  ~\cite{cesarone2011deep} &  ~\cite{de2011reliability}&   ~\cite{zhang2011survey} &  ~\cite{kaushal2016optical}  &   ~\cite{wu2018overview} &  ~\cite{biswas2018deep} & ~\cite{hall2020survey} &  ~\cite{konsgen2021current}   &This work \\
\hline
\multirow{6}{*}{Overview}     & Missions timeline      & $\times$     & $\times$   & $\times$     & $\times$        & $\times$ &$\checkmark$      &$\checkmark$     & $\times$   & $\checkmark$ & $\checkmark$       \\  \cline{2-12}
             & Demonstrations         &$\checkmark$         &$\checkmark$       &$\times$      & $\times$         &$\checkmark$    &$\checkmark$    &$\checkmark$    & $\checkmark$   &  $\checkmark$ &$\checkmark$  \\ \cline{2-12}
             & Existing systems         & $\times$   & $\checkmark$      & $\times$    & $\times$       &  $\times$   &$\checkmark$    &$\checkmark$     &$\checkmark$  & $\checkmark$   & $\checkmark$     \\ \cline{2-12}
            & Limitations     & $\checkmark$     & $\checkmark$    & $\checkmark$     & $\checkmark$        & $\checkmark$   &$\checkmark$     &$\checkmark$ &  $\checkmark$  & $\checkmark$  & $\checkmark$      \\ \cline{2-12}
            & Implementations       & $\times$    & $\times$     & $\checkmark$     & $\times$        & $\times$   &$\times$   &$\checkmark$    & $\checkmark$    & $\checkmark$    & $\checkmark$     \\ \cline{2-12}
             & Future perspectives      & $\checkmark$      & $\checkmark$   & $\checkmark$     & $\checkmark$        & $\times$ &$\checkmark$     &$\checkmark$  & $\times$    & $\checkmark$   & $\checkmark$      \\ \hline
\multirow{2}{*}{Basics}       & Link engineering         & $\checkmark$     & $\times$   & $\times$    & $\checkmark$   & $\times$  &$\times$    &$\times$   &   $\times$  &  $\times$  & $\checkmark$   \\ \cline{2-12}
           
             & Overall challenges      & $\checkmark$      &  $\checkmark$      &  $\times$    & $\checkmark$    & $\times$   &$\times$      &$\times$    & $\checkmark$    & $\times$     & $\checkmark$   \\ \hline
\multirow{3}{*}{Enabling technologies}  & Networking protocols      &$\times$    & $\times$   & $\checkmark$      & $\checkmark$     & $\checkmark$ &$\times$        &$\times$   &    $\times$  &  $\checkmark$    & $\checkmark$         \\ \cline{2-12}
                  & Modulation          & $\checkmark$      & $\checkmark$   & $\times$       & $\checkmark$     & $\checkmark$  &$\times$      &$\times$    &  $\checkmark$     & $\times$    & $\checkmark$         \\ \cline{2-12}
                 &  Coding              & $\checkmark$       & $\checkmark$    & $\checkmark$   & $\checkmark$      & $\checkmark$    &$\times$   &$\times$     &  $\times$   &  $\times$   & $\checkmark$        \\ \cline{2-12}                  & Receiver technologies   & $\checkmark$      & $\checkmark$    & $\times$     & $\checkmark$            & $\checkmark$     &$\times$        &$\checkmark$   &   $\checkmark$ &  $\checkmark$   & $\checkmark$   \\ \hline
Performance analysis  & Capacity        & $\checkmark$     & $\checkmark$      & $\times$     & $\times$       & $\times$  &$\times$    &$\times$     &  $\times$  &   $\times$    & $\checkmark$       \\ \hline
\multirow{4}{*}{Advances}            & OAM               & $\times$    & $\times$     & $\times$       & $\times$        & $\checkmark$  &$\times$    &$\times$    &   $\times$   &  $\times$   & $\checkmark$        \\ \cline{2-12}
                   & IPN              &$\times$        & $\times$   & $\times$          & $\checkmark$         & $\times$ &$\times$     &$\times$      &  $\times$  & $\times$     & $\checkmark$         \\ \cline{2-12}
                   & QC         & $\times$        & $\times$    & $\times$       & $\times$      & $\times$   &$\times$    &$\times$ &   $\times$      & $\times$        & $\checkmark$  \\ \cline{2-12}
                    & RF/FSO         & $\times$        & $\checkmark$    & $\times$       & $\times$      & $\checkmark$   &$\checkmark$    &$\checkmark$     &  $\times$  &    $\times$ & $\checkmark$  \\ \hline

\multicolumn{2}{|c|}{DS context }              & Fully   & Fully    & Fully & Fully  & Partially    & Fully    & Fully &  Partially & Fully & Fully  \\ \hline

\multicolumn{2}{|c|}{System  technologies}           & FS0  & RF\&FSO  &  RF\&FSO  & RF       & FSO   & FSO       & FSO  &  RF\&FSO   & RF\&FSO   & RF\&FSO       \\ \hline

\multicolumn{2}{|c|}{Time relevance}             & 2011   & 2011   & 2011   & 2011       & 2017    & 2018       & 2018  &    2020 & 2021  & 2024       \\ \hline

\end{tabular} 
\end{table*} 
 
 \subsection{Organization}
 The remainder of the paper is structured as follows.
 Section~\ref{sec:dscomb} outlines DSOC advantages  over RF and implementation efforts and challenges. Section~\ref{sec:dsPMod} analyzes 
 DS links physical characteristics and   discusses  related research directions. While Section~\ref{sec:dsretech2} characterizes DSOC  capacity limits, Section~\ref{sec:dscomd} surveys  DS  networking protocols and  related research efforts. Section~\ref{sec:dsAdvances} investigates on  
 {the} state-of-the-art and challenges of DSOC  advances including, IPN, OAM, \acrshort{qc}, and  \acrshort{rf}/FSO. 
Finally, Section~\ref{sec:conc} concludes the survey. 

\section{DSOC Advantages and Challenges}
\label{sec:dscomb}
Optical communications promise tens to hundreds of times increase in data rates, thanks to the high-available bandwidth in the license-free spectrum in which the technology operates~\cite{kaushal2016optical,khalighi2014survey}. Such a capacity improvement has already been proven in demonstrations~\cite{cornwell2017nasa}. 
To illustrate further the gain of using {optical communications} compared to RF in DS, based on~\cite{fielhauer2012concurrent}, the figure of merit (\acrshort{fom}) metric can be defined as follows:
\begin{equation}
\text{FoM} \delequal {10}\times \log \left( {\frac{D^{2}\times{R}}{{D_{r}}\times{P_t}}} \right), 
\end{equation}
where $D$ is the end-to-end DS link distance in the astronomical unit (\acrshort{au}). 
Here, $R$, $P_t$, and $D_{r}$ are the data rate in bits per second, transmit power in Watts, and receiver diameter size in meters, respectively. 
As it accounts for a space terminal performance  measured through data rate and cost  reflected through receiver aperture size and communications power, the FoM metric provides a direct comparison between the RF and FSO technologies at the system level  at any distance.

\begin{figure}[t!]
\centering
\includegraphics[width=1.0\linewidth]{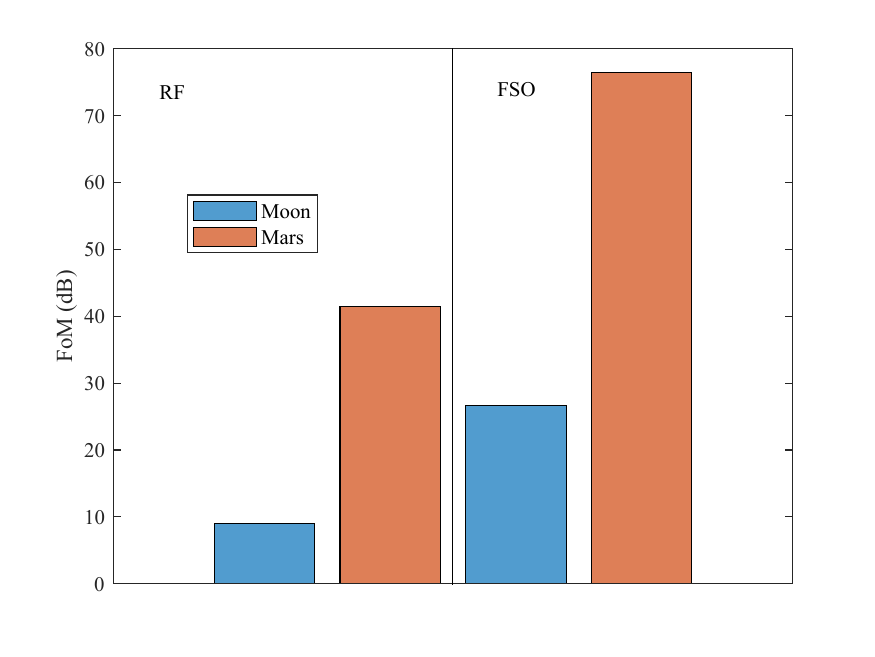}
\caption{FoM of DS communications systems.}
\label{Fig.ComRfFSO}
\end{figure}

Based on the projections available in~\cite{arapoglou2017benchmarking,wu2018overview,cornwell2017nasa}, Fig.~\ref{Fig.ComRfFSO}  demonstrates FoM for  Lunar and Mars. 
Unsurprisingly, a Lunar FSO system represents more than 15 dB gain over its RF counterpart. 
The improvement seems to be even more substantial over Mars distances. 
However, implementing FSO in DS communications faces a number of unique challenges.

Below, we discuss different challenges facing DS communications at the link level and the DSOC link budget along with the challenges in optical communications.

\subsection{DS Communications Challenges}
Ensuring high-capacity and reliable  communications in DS faces the  challenges listed below. 
\begin{itemize}
\item  Ultra-long, which based on the  International Telecommunications Union exceeds 2  million kilometers,  and time-changing distances between Earth and probes (satellites in other planets' orbiters, rovers, etc.). 
\item Interference from communications with other missions,  radiation from the Sun, other celestial bodies, and cloud cover.
\item Solar scintillation, atmospheric conditions, and other channel impairments.
\item Variability of  Sun-Earth-probe angles and the effect of that on blocking line-of-sight links. 
\item The  burden of size, weight, and power (SWaP) constraints of spacecraft  systems. 
\item The readiness and availability of transmitting and receiving technologies and communications protocols. 
\end{itemize}

The communications with targets in DS require sending signals over very long distances, leading to strong power dissipation and hence huge signal attenuation.
Even with relatively large-gain transmit and receive antennas, the received power can still be very weak.  
Also, the long-distance links result in an inevitable significant latency that prevents real time communications and control.
For example, a two-way communications with  Mars may undergo 40 minutes transmission delay.  

Furthermore, the Solar flares are shown to make {communications} signals  weaker and noisier~\cite{hemmati2011deep}. Besides, Solar scintillation and turbulence  lead to performance degradation. It is important to carefully account for the near-Sun effect on the signal strength to compensate for any losses and prevent any damage to  probe {systems}' circuits.
Also, due to the movement of Earth and probes, the Sun may completely block  communications path. 

In addition, \acrshort{swap} constraints are {of} major concerns  
not only for efficient spacecraft design but also for communications performance improvement. Designing  a space mission
with ultra-lightweight resources  is thought of as a way to utilize a better communications performance~\cite{hemmati2011deep}. There are technologies such as Gossamer that has the premise to design spacecraft structures with thin, low-modulus, and low-mass materials. An example of Gossamer technology application is the inflatable antennas; which could have an areal density of a fraction of a kilogram per square meter~\cite{chmielewski2005gossamer}. Also, in the direction of optical communications, the development of ultra-lightweight telescopes is realized; an example of which is the Cassegrain telescope which has composite mirrors with areal density in the range of 1\textendash 10 kg/m\textsuperscript{2}~\cite{boone2004development}.

{Lastly, existing terrestrial technologies, especially at the physical and protocol levels, can not be adopted directly in DS. For example, 
 as they are simpler in implementation, less lossy, and more power and bandwidth-efficient, digital modulation schemes are preferred in DS RF-based communications.
Nevertheless,  amplitude-based modulations that can not be generated with nonlinear amplifiers are not suitable  for DS as working in saturation mode is needed to maximize power efficiency~\cite{simon2005bandwidth}. 
Also, in networking, commercial terrestrial network architectures, for example transmission control protocol (TCP)/internet protocol (IP), are not efficient to ensure data delivery in the DS time changing  environments~\cite{burleigh2003delay}.}

Advanced techniques are needed to overcome  DS challenges and account for unique  link characteristics which we elaborate on next. 
%
%
 
\subsection{Link Engineering of DSOC}
\label{sec:dscomc}

Designing DS optical links needs to account for all link parameters and communicating entities' characteristics. This includes the end-to-end distance between Earth and final targets, mission requirements, required capacity, atmospheric conditions,  transmitter-Sun-receiver angle, etc.
The received power at the target after the receiver telescope and before the photo-detector can be expressed as~\cite{lyras2019deep}:   
\begin{equation}
\label{RecPowequation} 
{{P}}_{tar}{=\ }{{P}}_{{t}}\;{{G}}_{{t}}\;{{G}}_{{r}}\;{{L}}_{{fs}}\;{{L}}_{{pt}{ }}\;{{L}}_{{a}}\;{{L}}_{{c}}\;{{L}}_{{s}}\;{{\textrm{$\eta$}}}_{{t}}\;{{\textrm{$\eta$}}}_{{r}},
\end{equation}
where: 
\begin{itemize}
    \item $P_t$  is the transmitted power in Watts.
    \item $G_t$ and $G_r$ are the transmitter and the  receiver aperture gains, respectively~\cite{klein1974optical, booth2018optical}. 
    \item $L_{fs}$ is the free space loss factor. Being {inversely} proportional to the link distance squared, $L_{fs}$ represents the link dominant loss factor.
     \item $L_{pt}$ is the pointing loss factor. A pointing error of $\lambda/D_{t}$ radians, where $\lambda$  is the signal wavelength and  $D_{t}$ is the transmitter diameter, corresponds to approximately  85\%   intensity reduction~\cite{hemmati2020near,klein1974optical}.
    \item $L_a$ is the atmospheric transmittance due to the aerosol and molecular particles in the atmosphere that cause the absorption and scattering of the signal. The recommendation is to operate close to the 1550 nm wavelength {range} for low attenuation transmission~\cite{hemmati2006deep,alkholidi2014free,CCSDS}. 
    \item $L_c$ is the transmittance of the semi-transparent ice clouds, cirrus, present in the communications path\cite{degnan1993millimeter}.
    \item $L_s$ is the scintillation loss factor, which refers to the random optical power fluctuations caused by the atmospheric turbulence (AT)~\cite{osche2002optical,giggenbach2008fading}.
    \item $\eta_t$ and $\eta_r$ are the transmitter and receiver efficiencies, respectively~\cite{moision2003downlink}.
\end{itemize}

    The various DSOC impairments reflect the importance of  {the} link engineering  in deriving the needed transmit power to meet a certain detection sensitivity and hence a service quality. 
    It is worth mentioning  that the atmospheric and weather limitations are not of concern for optical space-to-space {links~\cite{hemmati2011deep,kaushal2016optical,hurd2006exo,djordjevic2010ldpc,TrichiliJOSAB20},} making the choice of the FSO
   over the RF in these links of great value. This fact can be reflected in the drop of  $L_a$, $L_s$, and $L_c$ loss terms in~\eqref{RecPowequation}.     

    More investigations about {the} potentials of the FSO technology in improving DS communications along with challenges facing its implementation  are provided below.
\subsection{DSOC Promises and Challenges}
\label{sec:dscome}
 RF technology is currently dominating space communications facilities.
 The RF systems' maturity and  the cost of implementing FSO are  delaying the technological migration.
 The  nonobligatory  of RF  line-of-sight  link availability~\cite{mukherjee2016wireless} along with its {tolerance to} atmospheric effects and weather conditions~\cite{kaur2012performance} {are  also} contributing factors to  deferring  the transition to FSO. 

 Nonetheless, the fairly limited  RF bandwidth requires licensing. Furthermore,    RF communications are susceptible to eavesdropping~\cite{mukherjee2016wireless}. 
 In addition, having the demand for high-data rates  comes at the cost of increasing SWaP requirements which can be unfeasible  and hence restrict RF system performance  upgrades. 
 {For instance, if we are acquiring several gigabits per second as downlink data rate from the closest distance to Mars, according to~\cite{cesarone2011deep}, we would need the Mars spacecraft to have a 5 m antenna with a {transmitting power {of} around 200 W}, and a multiple of 34 m  antennas in an array as the  ground Earth receiver. In contrast, for optical, the Mars spacecraft would have a 1 m aperture and a transmitting power of 50 W, and the receiver telescope aperture size would be in the range of 10 m.}

FSO, on the other hand,  is emerging  to address the shortcomings of conventional RF systems. 
  FSO beams, as they expand thousands {of} times slower than RF~\cite{franz2000optical},  demand significantly lower SWaP  allowing more flexible system upgrades and  secure communications~\cite{mukherjee2016wireless}. Not to mention the almost limitless bandwidth with no license requirements the FSO offers~\cite{hassan2020free}. {Nonetheless, achieving high  performance over DS {distances} is challenging~\cite{hemmati2011deep}.} 
A high gain is required to make up  for the very long distances loss  along with other losses discussed in  {the} previous subsection. 
 DSOC implementation hurdles can be compensated through several  approaches, that come with their  challenges, as we discuss below.

\begin{itemize}
    \item Increasing the power efficiency by relying on laser beams with large-gain transmitters and large-aperture diameters. Nonetheless, the cost of building large optical ground transmitters is considered to be  an implementation concern~\cite{deutsch2019creating}. Especially  {since} the average power requirement for a DS link typically would be in the range of 1\textendash 20 W and peak power would be  ranging from  100 to 1k W~\cite{hemmati2011deep}.
    
    \item Accurately tracking the space terminals using beacons to point lasers towards them before sending data. 
     {A} narrow beam requires  accurate pointing in order for the signal to be received at the destination, which is challenging considering DS distances~\cite{oaida2014optical}. 
    Particularly, regarding flight systems, high accuracy of the beam pointing is needed, especially for downlink signals from DS that are typically with a beam-width in the range of 2\textendash8 $\mu$rad. Adequate beam acquisition at destinations can be achieved at the cost of computational and hardware complexities though~\cite{bashir2020signal}. In addition, tracking and pointing beams toward mobile receivers in space require   efficient design of accurate tracking algorithms. 
    
    \item Relying on large-gain receivers. The received power to spacecrafts would be around or less than 10\textsuperscript{-10}W~\cite{lyras2019deep}. As a consequence, that requires  high gain receiver telescopes, 
    which can  make it a challenge to obtain a functional yet optimized design of the spacecraft resources~\cite{hemmati2011deep}.  

    \item Considering advanced signaling and detection techniques that maximize photon efficiency, i.e., {the} number of bits per photon. Achieving this  would  also include getting high-detection sensitivity and acquiring efficient modulation and  coding schemes~\cite{biswas2017status}. 
    
    \item {Merging FSO with other  technologies, including OAM for example, to get the most out of its high capacity and further enhance the quality of DS data returns~\cite{kaushal2016optical}. More investments though need to be put forward to facilitate such a {merge}.}
\end{itemize}

 {There are several efforts being brought forth  to facilitate the full implementation of the FSO  technology in DS~\cite{wilson1997results,biswas2006mars,cesarone2011deep,grein2015optical,oh2017development,biswas2018deep,karmous2023connecting} and address the DS FSO technology engineering challenges  and   maturity requirements~\cite{hemmati2011deep,fielhauer2012concurrent}.
 In this paper, we investigate the efforts {made} in DSOC and identify  open research directions.}   

\section{DSOC Physical Layer Key Characteristics}
\label{sec:dsPMod}

This section reviews the different physical  techniques  that mitigate a number of the previously discussed challenges to ensure DSOC  reliability.   
\subsection{Modulation Schemes for DSOC}
\label{sec:dsPMod1}
Conventional RF-based space communications rely commonly on binary phase shift keying (BPSK) modulation~\cite{statman2004coding}, where a carrier modulates data using two phases that are 180 degrees apart. \acrshort{bpsk} has been favored in such an environment, thanks to {its} low bit error rate (BER) in the low signal-to-noise (\acrshort{snr}) ratio regime.
However, for DSOC, there are other direct detection-based modulation schemes of interest including on-off keying (OOK), pulse position modulation (PPM), wavelength modulation, and pulse intensity modulation.
\acrshort{ook} and \acrshort{ppm} are the most recommended among the methods mentioned above, thanks to their simplicity and  power efficiency {in photon-starved regimes}~\cite{kaushal2016optical,hemmati2011deep}.

 As PPM is more power efficient, it is 
 more preferred than OOK for DSOC. On the other hand, OOK is more suitable for near-Earth links as it is {bandwidth-efficient and cost-effective}~\cite{kaushal2016optical}.
 While, OOK modulates one as a pulse and zero {in} the absence of a signal, M-ary PPM modulation, which is often used with photon-counting (PC) detector~\cite{hemmati2011deep,hemmati2006deep,li2012dual}, works by  dividing each channel symbol period into $M$
time slots. $\log_{2}M $ bits are modulated by transmitting a single pulse in one of the $M$ possible time slots of a symbol.
The information in the PPM is encoded in the position of the pulse, making it robust against noise. PPM requires a high bandwidth {though}~\cite{moision2005coded}. The good news, however, is that the high bandwidth is readily offered by FSO and hence it is not of much concern.

To characterize the effectiveness of PPM for DS applications, we provide the capacity offered by the PPM-PC receiver in the presence of noise and using a Poisson channel model, typically used for DS system characterization~\cite{kakarla2020one}. The approximate system capacity, denoted by $C_{PCR}$, is given  by~\cite{tyson1996adaptive}
\begin{equation}\label{Capacityequation}
C_{PCR}= \frac{1}{\mathrm{{ln}(2) }{E}} \Bigl(\frac{{P_r}^2}{P_n\frac{2}{M-1}+\frac{P_r}{\mathrm{ln}(M)}+{P_r}^2\ \frac{M\ T_{slot}}{{\mathrm{ln}(M)\ }{ E}}}\Bigr),
\end{equation}
where the photon energy  $E =(hc)/ \lambda $, $h$ is Planck's constant, $c$ is {the} light velocity, and $\lambda$ is the signal wavelength. 
$M$ is the modulation order. $P_r$ and $P_n$ are the received and noise power, respectively. $T_{slot}$ is the pulse duration.
The three terms in the denominator of the second term   of~\eqref{Capacityequation} correspond to three different behaviors when dominating.
As $P_{n}\frac{2}{M-1}$, which is a noise-limited term, dominates, the capacity scales as a quadratic function of the signal power. 
 On the other hand,  $P_{r}\frac{1}{\ln(M)}$, which is the quantum-limited term, leads to a linear scaling of the capacity. Finally, the  bandwidth-limited  term, ${P_{r}^{2}}\frac{M T_{slot}}{\ln(M)E}$ leads to capacity saturation. 
Note that when the received $P_r$ is sufficiently high, the system capacity can be approximated as 
\begin{equation}\label{Capacityequation1}
C_{PCR}\approx \frac{1}{\ln(2)}\frac{\ln(M)}{M\ T_{slot}}.
\end{equation}
 If the received power is sufficiently high, one can easily  observe that it is best to use a low PPM modulation order.  

\begin{figure}[htbp]
    \centering
    \begin{subfigure}[t]{0.5\textwidth}
    \hspace*{-0.9cm}
     \centering
    \includegraphics [width=\textwidth]{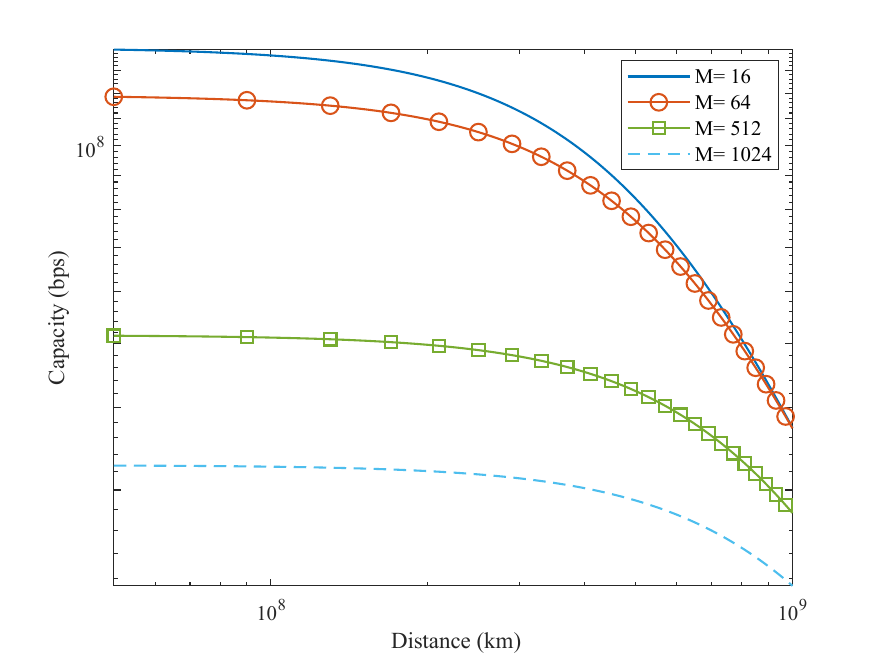}
    \caption{Mars Distances Capacity}
   \label{A:CapacityVSDistance1}
    \end{subfigure}
  \hfill
      \begin{subfigure}[t]{0.5\textwidth}
      \hspace*{-0.9cm}
       \centering
    \includegraphics[width=\textwidth]{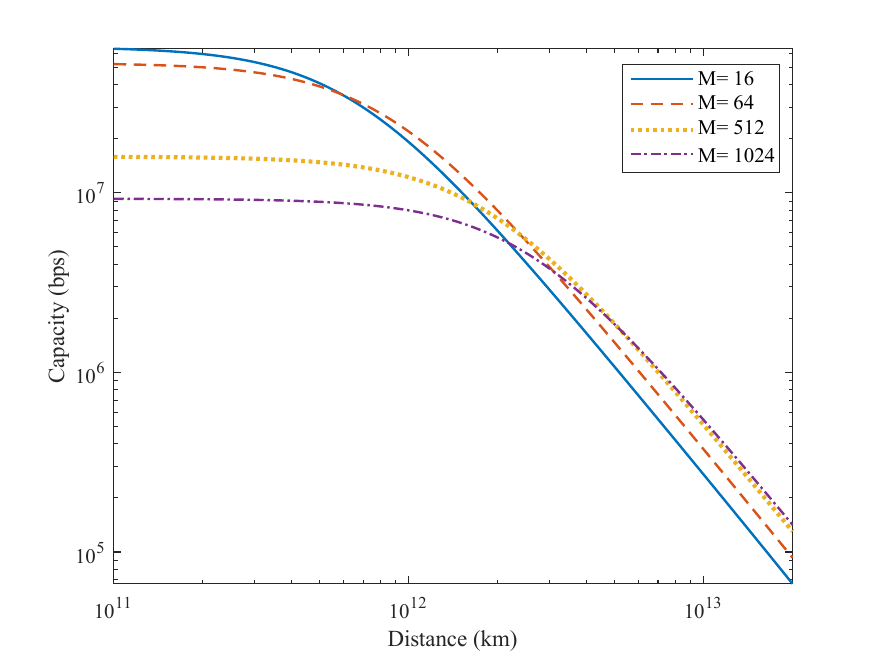}
   \caption{Beyond Mars Capacity}
    \label{B:CapacityVSDistance2}
    \end{subfigure}
    \caption{Link capacity versus distance.}
  \label{CapacityVSDistance}
\end{figure}

In Fig.~\ref{CapacityVSDistance}(a) and~\ref{CapacityVSDistance}(b), we show the PPM systems achieved capacity  for Mars and beyond Mars distances. Here we pick the antenna diameter for  {the} transmitter as 0.22 m and the receiver as  4 m~\cite{lyras2019deep,deutsch2019creating,biswas2017status}. 
 The figure shows that as the modulation order decreases, the capacity improves significantly but only up to a certain distance. 
 That is intuitive as increasing the modulation order implies a reduction in the transmission rate and \acrshort{ber}. As depicted in Fig.~\ref{CapacityVSDistance}(b), {as a  signal travels a longer distance, the trend changes. I.e.,
 as the distance gets larger and larger, to maintain a certain BER performance, the modulation order may need to be increased.}
 Fig.~\ref{CapacityVSDistance} indicates   how crucial the increase in DS communications distance is in determining the PPM system parameters and hence performance, as  discussed in~\cite{karmous2023connecting} and several references therein. 
 The impact of combining the PPM modulation technique with  coding is also of crucial importance as we will discuss.  

\subsection{Coding for DSOC Links}
 \label{sec:dsPMod2}
Channel coding is a very important block in the transmission chain to enhance performance and ensure   reliability. 
Applying ideal error correction coding (\acrshort{ecc}) may lead to meeting a link capacity limit,  {the}  maximum possible achievable capacity.
However, such ideal codes are hard to construct in practice. Therefore, it is important to consider code efficiency, which refers to the difference between the capacity limit and the error code threshold at a specific BER,   when designing  communications links~\cite{Moision2012Designtool}. 

In the early missions to DS,   {in the 1960s}, a decision was needed on whether to rely on a coding method or the other, namely  block codes or convolutional codes. 
During the design of  Mariner '69  Mars mission, for example, block codes were the best options given that {the} developed sequential decoding algorithms were not that efficient for DS.  
Such a choice led to a poor data rate, though. 
Thanks to the development of a more efficient sequential decoding algorithm,  convolutional decoding methods, for example, Turbo codes  have been adopted {for} various communications channels  including DS~\cite{divsalar1995multiple}, starting from Pioneer 9 mission~\cite{massey1992deep}   which was designed after Mariner '69 but launched  earlier. {Such a change offered a  coding gain, which is the power gain over the uncoded case, of 3 dB~\cite{lin1983error} compared to that of 2.2 dB~\cite{massey1992deep} in the case of  {the} block code used by Mariner '69.} 

\begin{figure}[t!]
\centering
\includegraphics[width=0.85\linewidth]{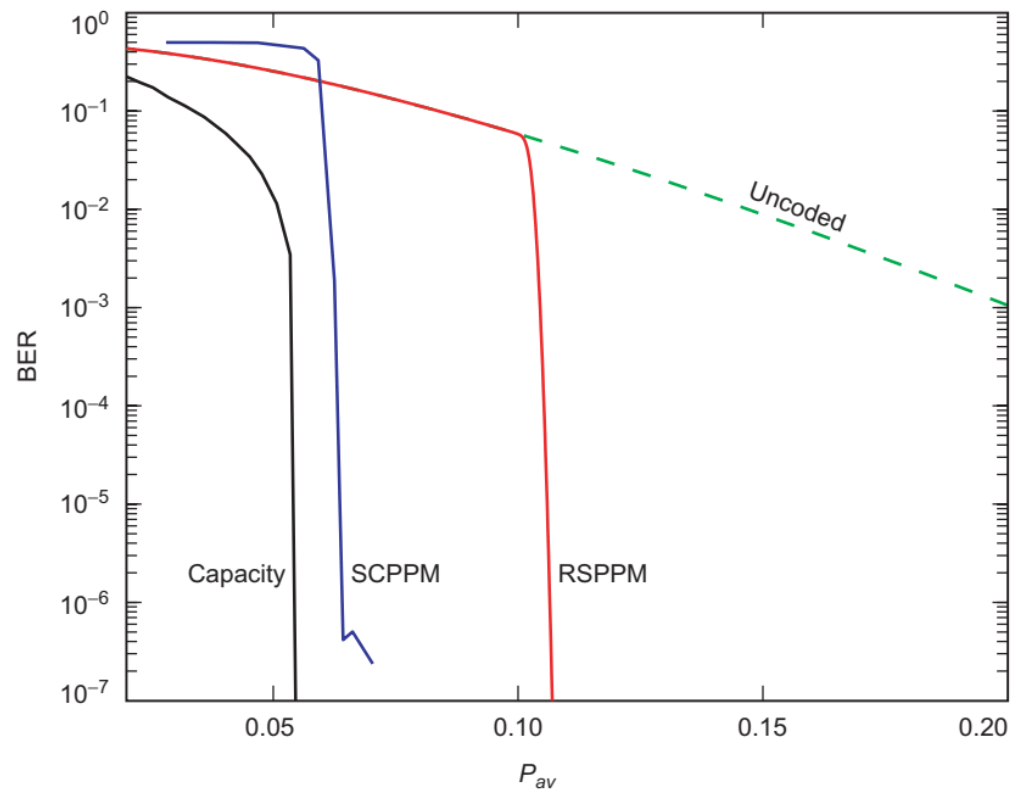}
\caption{Coding techniques BER compared to uncoded communications and capacity limit~\cite[Fig.~4-35]{hemmati2006deep}.}
\label{Fig.coding}
\end{figure}

 Reed-Solomon (\acrshort{rs}) codes are commonly used with convolutional codes for DS applications as they are efficient in correcting complex and bursty errors~\cite{de2011reliability,hemmati2006deep,hemmati2011deep,mceliece1994reed}, and have  in the order of 2.5 to 3 dB efficiency~\cite{hemmati2020near}. As a reference,  an uncoded PPM-based system would have a code efficiency in the order of 5 dB~\cite{moision2005coded}. Nonetheless, the most recommended and efficient DS coding scheme is the serially concatenated PPM (\acrshort{scppm})~\cite{cheng2006sat05,SCPPM4DS2006,moision2005coded}. This coding technique has a code efficiency between 0.5 and 1 dB~\cite{hemmati2020near,moision2005coded}. Fig.~\ref{Fig.coding} shows the BER of SCPPM versus average signal power measured in single photons detected per modulation order and {denoted by} {$P_{av}$},  compared to the capacity limit, Reed-Solomon  PPM  (\acrshort{rsppm}), and uncoded PPM. The figure  confirms the superiority of SCPPM over RSPPM, as  it   clearly {demonstrates}   lower coding efficiency  and  higher  gain. 

Recently, there have  been other  codes suggested for DS. For example, Divsalar et al.~\cite{divsalar2020wavelength} show that coded optical wavelength division multiple access (WDMA) with PPM can be a good candidate. 
Using the coding gain metric,  \acrshort{wdma} has a 7 dB gain for a 0.8 coding rate and a 12 dB coding gain for a coding rate of 1/6 at a frame error rate  (\acrshort{fer}) of 10\textsuperscript{-6}~\cite{divsalar2020wavelength}.

Since fixed-rate codes like low-density parity check (LDPC) and Turbo codes are not efficient for DS communications as  they fail to adapt to  changes  incurring in communications links due to time variability, Liang et al.~\cite{liang2020raptor} proposed the use of rate-less codes such as raptor codes, and hence achieve a better performance than \acrshort{ldpc} and Turbo codes. The authors propose to further combine spinal and polar coding to achieve even better performance.  
Further improvement can be offered  through the use of efficient detection methods  as we will elaborate {on} next. 

\subsection{Detection Techniques for DSOC}
\label{sec:dsretech}
Advances in detection technology play an important  role in providing reliable DSOC. 
As indicated in Section~\ref{sec:dscomb},  having improved sensitivity and efficient detection are regarded as means to {overcome faint DS  {signals} issues and design limitations}.
Detection receivers can either be coherent or non-coherent. 
Coherent  receivers can have a higher spectral and dimensional information efficiency in one state system, examples of which are  heterodyne and homodyne receivers. 
Dimensional information efficiency,  according to~\cite{dolinar2011approaching}, is defined as the efficiency  of information transfer  per  unit  dimension or  optical communications degree of freedom.  
Even though coherent receivers suffer from limited photon efficiency, they tend to be more practical when very high data rates are required.  
Non-coherent receivers, on the other hand, are energy efficient but come at the cost of  exponentially decreasing dimensional efficiency~\cite{dolinar2011approaching}. 
Rendering to their energy efficiency, non-coherent receivers which are alternatively called direct detectors are commonly used in DS~\cite{hemmati2020near,caplan2007laser}.
There are different types of direct detectors among which are avalanche photodiode detectors (APD), PIN (p-type, intrinsic, and n-type) diodes,  and \acrshort{pc} detectors.

\begin{table*}[htbp]

\caption{DSOC DETECTORS}
\begin{center}
 \begin{tabular}{|P{1.5cm}|c|c|P{3cm}|P{1.5cm}|P{1.5cm}|P{2cm}|P{3cm}|} 
 \hline

\multicolumn{2}{|c|}{ \Centering Detectors/Receivers}    & \Centering Type & \Centering Implementation & \Centering Sensitivity  (\acrshort{ppb}) & \Centering Data rate (Mbps)   & \Centering Advantages & \Centering Disadvantages \\ \hline
     InGaAs  & \multirow{3}{*}[-14ex]{\rotatebox[origin=c]{90}{Direct}}  & PIN     &   Cubesats for DSOC systems &     -     &   $163$    & Relatively high-temperature operation  &      Relatively  high received power requirement, not a good option as a ground receiver due to atmospheric losses \\ \cline{1-1}\cline{3-8}
      GmAPD   &            &      \acrshort{apd}       &  Mars laser communications demonstration (\acrshort{mlcd}) project, imaging system for DSOC Project&       2     &   $34.9$      &  No read out noise penalty  &  Limited bandwidth  \\ \cline{1-1}\cline{3-8}
        SNSPD  &            &      PC         & Lunar laser communications demonstration (\acrshort{llcd}) demonstrations, DSOC Project&    0.5     &     $781$    &     High-detection efficiency ($98\%$)   &   {Requirement for cooling down to 2\textendash4 K,  high photon rate detection (multiple of Gbps) can be hard to attain}
        \\\hline
           SQ   & \multirow{3}{*}[-7ex]{  \rotatebox[origin=c]{90}{ Coherent}} &  Homodyne   &  Near-Earth and DS applications &  1.5   &     156   & Energy efficient       &  Limited bandwidth \\ \cline{1-1}\cline{3-8}
             SPC &           & Heterodyne & EDRS, DS communications (suggested) & 10 &  1800 &   Robust to severe background noise, near-Sun communications &   High-bandwidth requirement  \\ \cline{1-1}\cline{3-8}                    

             PSA   &              &  Pre-amplified & DS application  demonstrations &  1    &    10500   &   Noise-free amplification     &    High performance in a limited spectral efficiency range \\ \hline
             
\end{tabular}
\end{center}
\label{tab:dsoc_det1}

\end{table*}

Due to their high sensitivity compared to other conventional direct detection methods, PC detectors are gaining great attention and are recommended along with PPM modulation to be used in the future DSOC.
These detectors can count a single photon incident on them. PC detectors can be viewed as an extension of APD detectors with infinite gain in which a digital output signal is generated for each detected photon~\cite{caplan2007laser}. They can offer up to 20 dB gain along with PPM over APD and pre-amplified direct detectors~\cite{hemmati2011deep},  {and sensitivities of a few photons per bit~\cite{kakarla2020one}.} The PC detection process, though, is affected by noise sources which are  $i)$  background noise that comes from the diffused energy from the sky, planets, or stars in the field of view of the receiver, $ii)$ dark counts {of} electrons generated by the detector, and $iii)$ receiver thermal noise~\cite{lyras2019deep}. In addition to noise, the losses that degrade the PC detection efficiency are
$i)$ blocking loss which results from the fact that a detector gets blind after a detection event missing all incident photons until reset. Blocking loss limits a detector to {counting} at most one photon per reset time or dead time~\cite{moision2003downlink}, and  
 $ii)$ {jitter loss} which is the random delay that {elapses} from the incident of a photon on a detector to the time an electrical output pulse is generated in response to that photon. 
A significant loss occurs if the standard deviation of the jitter is in the order of the slot width of the pulse~\cite{moision2008Farrjitter}.
The choice of the detection and modulation schemes is essential in mitigating other losses resulting from link impairments. Detectors that handle a small duty cycle in the range of a few nanoseconds, for example, are resilient to long DSOC channel losses~\cite{book2019physical}. An example of a detector that can provide such a requirement is the superconducting nanowire single photon detector  (\acrshort{snspd})~\cite{stern2007fabrication,boroson2014overview,robinson2006781,huang2018high,shaw2018superconducting,JPL2021SuperNano}, which is the {highest-performing} commercialized  PC detector in the frequency range of 30\textendash 30,000 THz~\cite{JPL2021SuperNano}.
This detector can support slot widths down to 200 ps, {timing jitter less than 10 ps~\cite{korzh2020demonstration},
and up to 98\% detection efficiency~\cite{reddy2020superconducting,stern2007fabrication}.} There are some other examples of detectors and receivers that are significant in DS communications, such as indium gallium arsenide (\acrshort{ingaas})~\cite{rajguru2014laser}, Geiger-mode avalanche photodiode (\acrshort{gmapd})~\cite{mendenhall2007design,pestana2021evaluation,horkley2016optical}, single quadrature (\acrshort{sq})~\cite{stevens2008optical}, European data relay system (\acrshort{edrs}) with single photon coherent (\acrshort{sps}) heterodyne detection~\cite{heine2014european,reine2005encyclopedia}, 
and phase-sensitive optical amplifiers (\acrshort{psa})~\cite{kakarla2020one}. 

In Table~\ref{tab:dsoc_det1}, we include the main characteristics of the aforementioned detectors and receivers, including sensitivity measured in photon per information bit (PPB),  supported data rate, advantages, and shortcomings.
\subsection{Discussions and Outlook}
We stated earlier the different potential modulation, coding, and detection schemes for DS communications. {It is worth reiterating that yet experimenting   {with} existing schemes efficiency needs to be conducted.} DSOC systems are expected to operate at the speed of several tens of gigabits per second and beyond and hence will require a major improvement over existing receiver technology in terms of both data rate and sensitivity~\cite{hemmati2006deep}. Thanks to their energy efficiency and simplicity, direct detection systems are preferred over coherent systems. Nevertheless, coherent detection has recently started to get more interest due to its ability to increase the capacity of DS communications~\cite{kakarla2020one}.

While using SQ homodyne receivers, without a pre-amplifier, resulted in a sensitivity of 1.5 \acrshort{ppb}  at 156 Mbps~\cite{stevens2008optical}, demonstrations have shown that pre-amplified erbium-doped fiber coherent receivers can achieve a 0.51 PPB increase in sensitivity and a 10 Gbps rate~\cite{lavery2012realizing,geisler2013demonstration}.  
  {In addition, it is also worth mentioning that there is some focus on in-phase and quadrature modulation homodyne detectors in optical transmission systems for terrestrial as well as DS communications as they promise to achieve higher-spectral efficiency than other schemes~\cite{pottoo2020development}.} 

\section{DSOC Physical Layer Performance Analysis}
\label{sec:dsretech2}
To decide on a modulation type,  coding technique, and  detection approach, the channel capacity needs to be determined or at least approximated for a given system setup. 
\subsection{Capacity Limit }
\label{subsec:capacitylimit}
  Shannon capacity gives a channel the maximum error free data rate in classical information theory~\cite{shannon1949Noise}.
 {However, in optical communications, channels are best described as quantum channels, which are shown to accurately describe  capacity limit of a corresponding system. For extremely   power-limited communications, as is the case in DS,  the capacity limit is best characterized  by the Holevo theorem, which approximates the solution of the quantum channel capacity limit~\cite{banaszek2020quantum,holevo2012quantum}.}
 Based on the Holevo theorem~\cite{dolinar2011approaching}, the analytical closed-form expression of the quantum capacity, which asymptotically {expresses} the dimensional information efficiency, measured in bits/dimension, is given by
\begin{equation}
{C_{d}}^{Hol}=ec_{p}2^{-c_{p}},
\end{equation}
where $c_{p}$ is the photon information efficiency (\acrshort{pie}) {which is the amount of information that can be transmitted per photon, measured in bits/photon,} 
    and $e$ is the exponential constant. 
    In theory, the optimal coherent state modulation achieves the ultimate quantum limit, {which is the capacity limit that can possibly be approached by a system based on the Holevo limit approximation}.
    Still, practical sub-optimal coherent receivers are operating near the ultimate Holevo limit but in the low photon efficiency region as they hit a brick wall of {$1.44$ bits/photon} and hence they are not efficient for DS applications~\cite{dolinar2011approaching}.

In the case of PPM-PC receivers, the  capacity can be approximated with respect to the Holevo limit as~\cite{banaszek2020quantum}
\begin{equation}
    C_{Holevo} = 2.561 C_{PCR}, 
\end{equation}
where $C_{PCR}$ is given by~\eqref{Capacityequation}.
  {Dolinar et al.~\cite{dolinar2012fundamentals} proposed using coherent states with PC to get closer to the Holevo limit. This is challenging and not practical from an implementation perspective, though.}
  Quantum number state (\acrshort{ns}), {also known as  Fock state~\cite{leonhardt2010essential},  communications could be an alternative, where the quantum state   {with} a well-defined number of photons are shown to be able to reach the Holevo limit~\cite{dolinar2012fundamentals}.
This could be challenging in practice, again, since it requires high-channel transmissivity, which is the probability of transmitted NS photon being received at the detector. 
To further illustrate this, considering OOK modulation, the  single photon NS capacity at high {PIE} can be written in terms of the channel transmissivity, denoted by $\eta$, asymptotically as 
\begin{equation}
    {c_{d}}^{1NS} = {C_{d}}^{Hol}(2^{f(\eta)}),
\end{equation} 
where if ${\eta}<<1$, then $f(\eta) = \eta/e$ and when $\eta{\sim}1$, $f(\eta) = ({\eta}-1)^{(\eta-1)}/e^{1-{\eta}}$. In the case of PPM, we have 
\begin{equation}
{c_{d}}^{1NS} = {C_{d}}^{Hol}(\eta/e).
\end{equation} 
As shown in Fig.~\ref{HalevovsQSvsPPM}, PPM cannot reach the Holevo limit even when $\eta = 1$.
With a transmissivity of one,  NS along with OOK gets to the ultimate limit. As the quantum NS transmissivity  gets higher, the capacity gets closer to the Holevo limit in the case of OOK.
We also show in the same figure the capacity limit for PPM-PC, and quantum NS modulation, with OOK and PPM.

\begin{figure}[htbp]
\centering
\hspace*{-0.3cm}
\includegraphics[width=1.11\linewidth]{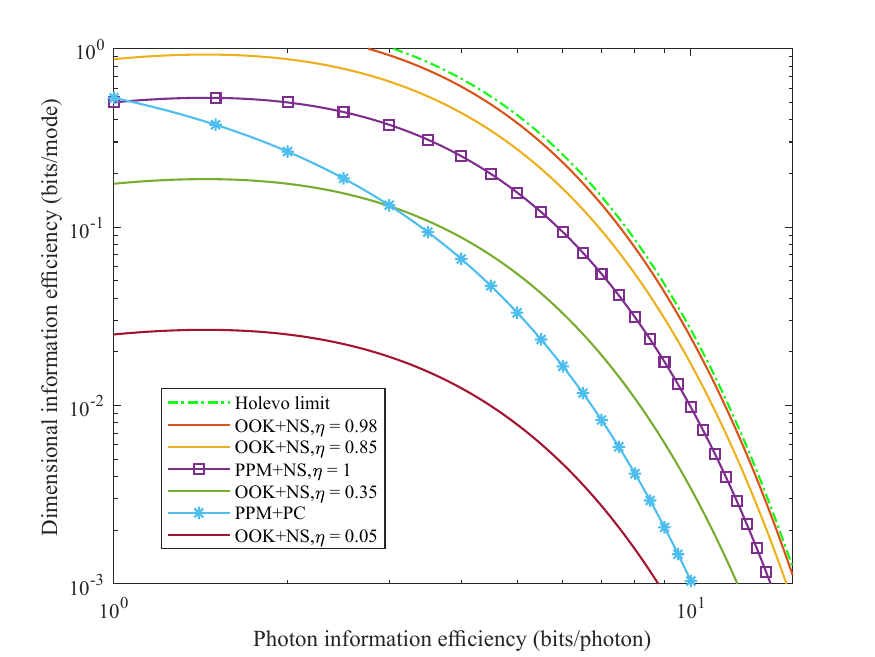}
\caption{Asymptotic capacity limit of different modulation schemes.}
\label{HalevovsQSvsPPM}
\end{figure}

 To analyze the DSOC link performance for a PPM-PC detector, in~Table~\ref{capacitynslot}, we show the soft decision capacity at various planets average distances. Soft decision capacity corresponds to implementing a soft decision receiver that passes on slot counts and uses them to determine the probability of  {sending a certain sequence of bits}~\cite{hemmati2006deep}.
Here we assume that the transmitter and the receiver aperture diameters are 0.22 m and 4 m, respectively, with a transmitted power of 4 W~\cite{lyras2019deep,deutsch2019creating,biswas2017status},     0.25 ns time slot,   PPM modulation order of 16, and   noise power  {of 1.1620$\times$10\textsuperscript{-16} W}.  The losses of Earth to Mars communications, for example in the considered setup, are as follows: a path loss of $~$366 dB, the sum of other losses for the received power not including the detection loss is 6.6 dB, and the detection loss is 4.35 dB.
We notice that the capacity dramatically gets smaller as we get to further planet distances. 

\begin{table}[!tbp]
\caption{DIFFERENT PLANETS CAPACITY}
\begin{center}
 \begin{tabular}{|c c c c c|} 
 \hline
 Planet &  Average & Average & Received &  Reached \\ 
        &  distance  & delay &   power  &  capacity  \\ 
        &    (km)    & (Minutes)& ($\mu$W)&  (Mbps)\\[0.5ex]
 \hline\hline
 Mercury & 58 million & 3.22  & 5.1856e-6 & 139.45 \\ 
 \hline
 Mars  & 225 million & 12.5 & 4.5053e-6 &  123.42 \\
 \hline
 Jupiter  & 778 million & 43.22 & 1.7741e-6 & 52.527 \\
 \hline
  Saturn & 1.2 billion & 66.66 & 9.2772e-7 &  28.173  \\
 \hline
   Neptune & 4.5 billion & 250 & 7.8951e-8 &  2.4598 \\ 
 \hline
  Pluto   &5.9 billion  & 327.77 &   4.6219e-8 & 1.4409  \\ [1ex] 
 \hline
\end{tabular}
\label{capacitynslot}
\end{center}
\end{table}

\subsection{Discussions and Outlooks}
Although direct detectors cannot reach  capacity limit, the capacity can be improved by other techniques including increasing  sensitivity and reducing the amount of background light detected. All relevant emerging technologies would be of interest as they can vitally define the future capabilities in DS environment. 
One of the innovative solutions that emerges to improve capacity limit and data rate, flexibility, scalability, and cost-effectiveness is compact array detectors~\cite{bashir2019free,lyras2019deep}. Array detectors decrease blocking and pointing losses and help with beam acquisition and tracking processes. 
It has been shown that, with the use of SNSPD, arranging a couple to tens of  detectors  significantly  reduces the blocking loss{,}   
 increases  efficiency to 93\%~\cite{shaw2015arrays} with up to 7.41 PPB sensitivity~\cite{hao2024compact}, and utilizes  data rates in the range of gigabits per second~\cite{dauler20061}.  
Such detectors hold  a continuous active area, ready to receive the upcoming photons, which could be as low as 3 $\mu$m
by 3.3 $\mu$m~\cite{dauler20061}. Although smaller active areas result in higher-data rates,  increasing  them improves  system detection efficiency.  
Hence, there seem to be efforts, see e.g.~\cite{xu2021superconducting,hao2024compact}, to get the benefits of relatively large-active areas  without sacrificing the data rate and facing the   reading out challenge which  results in increasing  temperature and complexity.
Despite the promises of array detectors, according to~\cite{vilnrotter2003optical,lyras2019deep}, in some cases the single receiver could outperform the array, for relatively long distances and in case of no AT.

There are a number of other technologies suggested regarding increasing the effectiveness of noise rejection, hence, improving PIE upper limit. That includes  the use of  nonlinear optical quantum pulse gating (\acrshort{qpg}) technique~\cite{jarzyna2024photon}, which can filter out the unwanted temporal noise.  
Even more improvements have been proven through the utilization of  photon number resolved (\acrshort{pnr}) detectors  which  interestingly, in contrast to conventional PCs for instance individual SNSPDs~\cite{sullivan2023photon},
are able to read more than one photon simultaneously. 
PNR detectors are highly regarded for quantum optic applications and quantum communications. 
 Making this type of detector along with other physical layer optimal solutions to yet be investigated  for further improvements in terms of capacity limit.

\section{DSOC Networking Protocols} 
\label{sec:dscomd}

In addition to the different physical layer techniques and their impact on DSOC performance as we  discussed in  
 the sections~\ref{sec:dsPMod} and~\ref{sec:dsretech2}, networking protocols are of paramount importance to make the communications links reliable and overcome a number of  different complex challenges facing DSOC  discussed in Section~\ref{sec:dscomb}. 
Seamless coordination between satellites, ground communications systems, and DS probes is needed. 
Matured terrestrial network protocols are  mostly {inefficient} in  DS environments where distances are extreme and connections are intermittent; hence, as suggested  in~\cite{hooke2001interplanetary},  calling for well-tailored network protocols is essential.
 NASA, for example, relies on delay/disruption tolerant networking (DTN) architecture~\cite{burleigh2003delay,CCSDSspaceprotocals,papastergiou2014delay} as an intuitive solution for space missions.  The architecture  can handle excessive delays and fast-changing environment while maintaining reliability.

The Consultative Committee for Space Data Systems (CCSDS), which is one of the most important organizations that discuss and set standards for space data systems, recommends architectures for space networks including DTN,  space communications protocol specifications (\acrshort{scps}) architecture~\cite{akyildiz2004state}, space packet protocol (\acrshort{sp})-based architecture~\cite{CCSDSspacepocket}, and \acrshort{ip} architecture. 
{{Rendering to} its ability to handle long and variable delays and link intermittent connectivity issues, \acrshort{dtn} is the most persuasive architecture.}
The automatic store-and-forward feature of DTN,  which allows packets to be forwarded to the next destination only if the availability of the corresponding link is ensured, 
 is making the DTN architecture a natural choice for severe environments~\cite{vasilakos2016delay,jones2006routing}.
\subsection{DTN Architecture Layers and Protocols}
DTN architecture consists of the application, bundle, transport, internet, data link, and physical layers. Below we discuss some of the fundamentals and functionalities of these  layers, along with corresponding protocols designed for DS internetworking.

 \textbf{Application layer:}
 In this layer, \acrshort{ccsds} file delivery protocol (CFDP) is a widely used protocol~\cite{kim2019performance}. In CFDP, the data units are delivered independently from each other, aiming to increase the amount of data~\cite{burleigh2002operating,haddow2014file}.  
Space missions, however, do not only rely on file transfers for all their communications requirements. CCSDS proposed {a} messaging exchange protocol called asynchronous message service (AMS) suitable  for various applications including real time control and robotic operation collaboration. \acrshort{ams} is designed to reduce the communications burden of space missions and hence their cost and risk of failure~\cite{book2011AMS}.

 \textbf{Bundle layer:}
Next to the application layer,  there is the bundle layer. This layer distinguishes DTN from the conventional TCP/IP architecture.
 {Due to frequent disruptions with long transmission delays, in a hop-by-hop transport fashion, the bundle layer that only has the bundle protocol (BP) adds a store-and-forward mechanism for each DTN node, i.e., every engine that runs BP in the network, and sends data in large packets called {bundles}.} 
 The BP mechanism is different {from} end-to-end protocols such as \acrshort{tcp}, {where} reliability is achieved through end-to-end acknowledgments and retransmission excluding the intermediate nodes~\cite{zhao2018network}. TCP or end-to-end protocols have a low transport efficiency if used in the DS network;   
 the hop-by-hop principle, {however,} enhances transport efficiency in  {such}  challenging environments~\cite{burleigh2003delay}. 
To enhance reliability and reduce congestion, \acrshort{bp} supports {custody transfer} which is an automatic retransmission request (\acrshort{arq})  mechanism that takes custody 
of a bundle, node-by-node, until it reaches its destination~\cite{book2015ccsds}. ARQ mechanisms are implemented in the  data link layer protocol as they are commonly used; however, in addition to that, they are
implemented in the upper layers of the DTN network architecture~\cite{zhao2018network}, to ensure the reliability of data delivery. An optional end-to-end acknowledgment feature~\cite{fall2003custody} could be used {in} situations where original data senders are hard to reach for a request of retransmission~\cite{araniti2015contact}. {Furthermore,  bundles are carried out  through convergence layer adapters (CLAs) which allow DTN to encompass different underlying protocols and hence work well in heterogeneous DS networks~\cite{fall2008dtn}.
Conventional routing protocols, most built for the TCP/IP architecture, are not efficient for time-variant DS networks~\cite{burleigh2003delay}.
  In efforts to tackle these challenges, contact graph routing (CGR) protocol has been developed for space networks {based on DTN architecture}~\cite{burleigh2008dynamic}. \acrshort{cgr} exploits the mission communications operation plans to construct a  {contact graph} that maps {time-varying network connections} which is used as a consequence for bundles' best route calculation. 

\begin{table*}[htbp]
\caption{{DTN ARCHITECTURE AND A CORRESPONDING CCSDS PROTOCOL STACK}}
\begin{center}
\begin{tabular}{| p{.07\textwidth} |c| p{.35\textwidth} |p{.2\textwidth} |p{.15\textwidth}|} 

 \hline
 \Centering Layers  & Protocol & \Centering Functions \& Mechanism  & \Centering Advantages & \Centering  Disadvantages \\ \hline
Application &  CFDP    & It is capable of file transfer,  delivery, and storage management services, namely loading, dumping, and controlling~\cite{book2020CCSDS}.
  
{It assumes that communications entities have storage mediums called {filestore}. The protocol operates by copying data between these mediums}. It has a delivery process that gives the capability of transferring files in arbitrary networks and works in a reliable as well as an unreliable mode 
& \RaggedRight Scalable, minimizes the resources required to only the necessary elements for file transfer operation~\cite{book2020CCSDS} & \RaggedRight Causes overhead in the network when used along with some other subsequent protocols~\cite{de2011reliability,kim2019performance}, does not  perform well for one-directional transmission long delay communications link~\cite{de2007performance} \\ \cline{2-5}
&  AMS  & AMS provides a asynchronous message exchange services~\cite{burleigh2006asynchronous}. It is designed to make the functioning parts of an application,  called {{modules}}, operate in isolation and make  communications between these modules self configuring. Hence complexity of the modular data system is reduced   & \RaggedRight  Simple to use,  highly automated, flexible, robust,  scalable, efficient~\cite{book2011AMS}     & - \\\hline
      
      Bundle     &  Bundle     &    It applies overlay network principle
      which  provides hop-by-hop transmission for data delivery and status reporting~\cite{book2015ccsds}. Information is sent in bundles, large groups of data, in a store-and-forward fashion with optional custody transfer~\cite{zhao2016performance}  &  \RaggedRight Data accountability, reliability in heterogeneous networks~\cite{book2015ccsds}    &  Relatively high-memory consumption~\cite{book2015ccsds}, not 100\% reliable~\cite{papastergiou2010does}  \\\hline

     Transport  &  \acrshort{ltp}     &  
   
      It provides optional reliable communications, as it  divides  blocks of data into red-part and green-part where the former needs to be delivered reliably but unnecessarily the latter~\cite{book2015LTP}    
     &  High throughput for the application level data, i.e., high good-put~\cite{wang2010licklider} & \RaggedRight In-order delivery is not supported, low performance for short distance and delay, with low BER link~\cite{wang2010licklider}

     \\ \hline

     Internet   &  IPoC   &
     \acrshort{ipoc} specifies the implementation of the IP  protocol over CCSDS SDLPs. The protocol data units (\acrshort{pdu}) of IP get encapsulated by \acrshort{ep} and then sent for framing by \acrshort{sdlp}~\cite{book2012ip}  &      IPv4 and IPv6 are commercially available technologies directly implemented in space, with good transfer performance  and cost effectiveness for short links~\cite{zhao2018network}
    &  \RaggedRight  Routing protocols fail in links with long delay that is caused by the very long distance~\cite{book2012ip}   \\ \hline
  
     Data link  &  USLP   & It utilizes features and functions of existing CCSDS data link protocols which are \acrshort{pro1}, TC, TM, and AOS, with added
     advances. It was developed to meet the requirements of space missions for efficient transfer of data~\cite{book2020USLP}
     
     &     Designed to reduce cost and complexity,  increases data rate, and meet the required data rate expected for DSOC, which can not be reached with the previous CCSDS standards~\cite{book2020USLP}   &  -     \\ \hline
     Physical  &  CCSDS-HPE     &  It specifies the characteristics of the transmitted laser signal~\cite{book2019space}   &     -    &  -  \\ \hline

\end{tabular}
\end{center}
\label{tab:dsoc_proto}

\end{table*}

 \textbf{Transport layer:}
TCP is considered an option for reliable data transmission over  {proximity} links in the transport layer. User datagram protocol (UDP) could be another option for loss-tolerant connections~\cite{sanchez2017support}. 
On the other hand, the Licklider transmission protocol (LTP)~\cite{book2015LTP} comes as an alternative transport protocol (\acrshort{tp}) designed to overcome long transmission delays and frequent interruptions. 
 {However, BP needs CLAs to interface between the different corresponding  TPs~\cite{caini2010dtn}.} Examples of such adapters are TCP, \acrshort{udp}, and LTP-based \acrshort{cla}s~\cite{hu2014memory}.  

LTP is the most common transport layer protocol used in DS.   It has most of the UDP and TCP functions; LTP, however,  does not perform handshakes or congestion control like TCP~\cite{sarkar2011survey}. Some efforts have been made to improve LTP in  memory consumption~\cite{zhao2015modeling}, reliability and energy efficiency~\cite{shi2020study}.  

 \textbf{Internet layer:}
When it comes to addressing nodes and packet delivery, i.e., internet layer functionalities,  nodes that are communicating in the same region,  protocols like IP can be used~\cite{book2007overview}.   In an attempt to make IP more compatible with the space environment, however, IP over CCSDS space links (IPoC)  is recommended~\cite{book2012ip}. 
If a sender is located in a  region different from the destination, however,  e.g. two different planets,  bundles are addressed differently for more effective routing. {Addressing schemes like IP in general indicate the geographic location of the host.  Such  methods, however,  provide no information about the host itself. In the case of interplanetary communications, the location of the host might change during the course of delivering the data, and hence  information about the host itself is beneficial for accurately assigning destinations~\cite{clare2010endpoint}.
A naming structure called universal resource identifiers (\acrshort{uri}), also referred to as endpoint identifiers (\acrshort{eid}),  {where} each DTN node gets associated with an identifier, a type of address that specifies the region, application, and host~\cite{fall2008dtn}, plays an important role  {in} achieving efficient interplanetary communications.}

 \textbf{Data link layer:}
 Packets from the internet layer get encapsulated either by space packets (\acrshort{sp})~\cite{CCSDSspacepocket} or encapsulation packets (EP) protocols~\cite{book2009encapsulation}.  
 Afterwards, packets go through the framing process in the data link layer with protocols provided by the space data link protocols (SDLPs). These protocols are telecommand (\acrshort{tc})~\cite{book2003tc},  telemetry (TM)~\cite{book2003tm}, proximity-1 (Prox-1)~\cite{proximity1space}, advanced orbiting systems (\acrshort{aos})~\cite{book2006space}, and unified space data link protocol (USLP)~\cite{book2018unified}. CCSDS SDLPs are responsible for transferring data with different types and variable lengths in space links. \acrshort{tm}, \acrshort{aos}, and \acrshort{uslp} can be used for optical communications~\cite{book2019space}.
Frames go through coding and synchronization data link sub-layer to prepare them {for} the physical layer, which transmits them to the medium based on the technical standard of the physical layer .

\textbf{Physical layer:}
For DSOC, high-photon efficiency (\acrshort{hpe}) specifications are recommended by CCSDS to be used in the physical layer. This standard characterizes the transmitted laser signal, including   modulation and related parameters, for space-ground and ground-space links~\cite{book2019space}.

\begin{figure*}[t]
\centering
\includegraphics[width=1\linewidth]{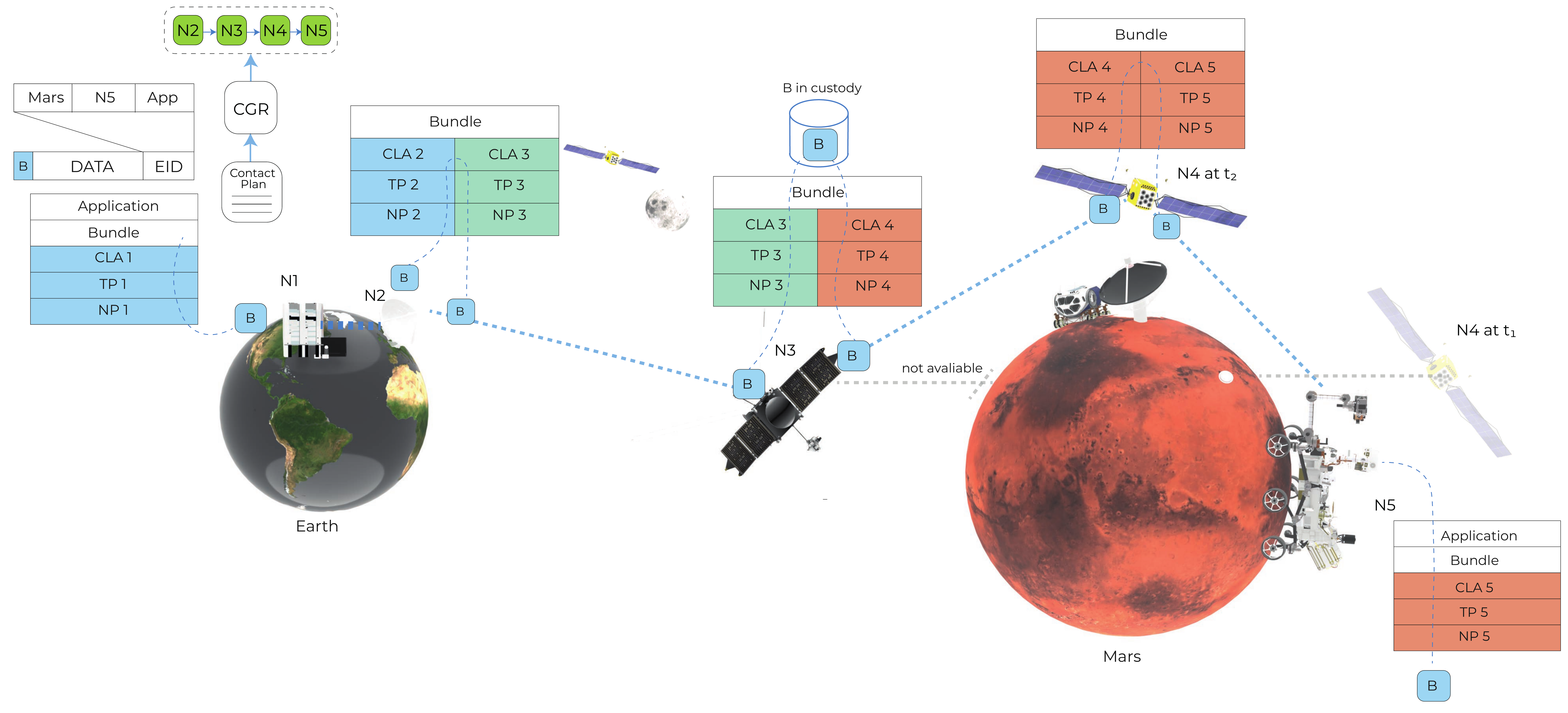}
\caption{BP custody transfer, CLAs, and  CGR dynamics visualization.}  
\label{NetworkprotocolsDTN}
\end{figure*}

In Table~\ref{tab:dsoc_proto}, we present some other aspects of the most significant and advanced DS protocols discussed above corresponding to their DTN layers. Additionally, to help visualize how DS DTN  features work, 
 Fig.~\ref{NetworkprotocolsDTN} further explains  the concepts of CLA, CGR, EID, and custody transfer. Let us build a scenario for transmitting
information  from  a mission control node  located on Earth N1 to a Mars rover N5. 
To have that happen, a bundle{,  denoted  by B,} is created  with the destination EID that includes the destination region and host in addition to the underlying application, App. 
A contact plan is also created giving  an estimation of future connections to nodes based on their locations and characteristics. The contact plan can be kept at N1 along with the routing table for centralized routing~\cite{fraire2021routing}.
The CGR algorithm at N1 uses the contact plan to estimate  possible routes  by evaluating parameters such as local time, 
delivery time,  bundle best transmission time, bundle priority, and bundle queue status.
In this scenario, it is assumed that  the best route to reach  N5  is through nodes N2,  N3, and then N4. As bundle B travels over networks, it  gets handled with CLAs  to integrate  different transport and other underlying protocols, if any. {At N3, B checks N4 which is assumed to be  unavailable at first contact, time t\textsubscript{1}. As a consequence, B is held in custody until the  node becomes, at time t\textsubscript{2},  accessible and hence B  finds its way to  reach the destination.}

\subsection{Discussions and Outlook}
When it comes to DTN protocols, it has been observed from the literature that some protocols in different layers may not work efficiently with one another. Hence, {a} lack of integration in the network stack may lead to internetworking issues. In this regard, although \acrshort{cfdp} accomplishes reliable file transmission, for example, it results in overhead when combined with BP and LTP.
 {The reliability of communications in DS is essential and important to ensure, but that comes  at the cost of overhead. 
  Kim et al.~\cite{kim2019performance} {attempted} to derive an optimization model to decrease the DTN network overhead for space missions caused by CFDP. 
However, investigating the performance of various protocols integration in the context of DS both in research and deployment is needed.  

    In addition to BP and its custody transfer scheme, which is the core of the DTN, several coding mechanisms have been adapted at the link layer. 
 To ensure reliability even further,  other coding techniques are needed  even in  {the} upper layers}. 
For example to handle the performance degradation of CFDP over long delays and one-directional communications links, erasure coding, which helps  in {overcoming} packet loss due to DS link degradation, is suggested to be implemented~\cite{de2007performance}. 
{Erasure coding can be used in addition to or instead of ARQ to improve data delivery as the latter  has been  shown to be inefficient, especially in long round trip links~\cite{hou2018application}, and  is known to increase processing complexity and storage utilization unlike erasure coding which speeds up  delivery~\cite{yu2013effect}. 
There are encouragements to conduct further research to optimize the use of erasure coding in the DTN architecture   and investigate other schemes to enhance overall DS communications performance. 
Even though the adoption of hop-by-hop transmission through BP  is inevitably required  {to handle} the intermittent availability of links while keeping throughput reasonable, 
end-to-end transmission  mechanism   may also be revisited along with other transport layer schemes to improve reliability avoiding  permanent bundle loss in case of a node failure.  
 One of these protocols is the deep space transport  protocol (\acrshort{dstp})  which serves fast file delivery, two times  faster than other conventional TPs under specific conditions~\cite{psaras2008ds}. 
However, when there is a  {difference in error rate estimation between a sender and a receiver}, DS-TP, as discussed in~\cite{papastergiou2009deep}, causes a malfunction in the retransmission process leading to inefficient use of bandwidth. 
Despite the efforts made to analyze this {protocol's} performance, much work is needed to fully understand its operation. Stream control transmission protocol
(\acrshort{sctp}) is another  TP, suggested  {in~\cite{fu2004sctp}},  {that, in certain conditions, 
surpasses conventional TCP} and can be utilized in orbiters that operate in the same region.  

Some other related protocols are also proposed e.g. the seamless IP diversity based network mobility (\acrshort{sinemo}) suggested, in~\cite{chowdhury2006sinemo,hossain2011cost},  to be adopted by mobile entities in space while addressing the drawbacks of the network mobility (\acrshort{nemo}) basic support protocol (\acrshort{bsp}), proposed to be used by NASA. 
These mobility protocols pose security risks, though, as in some cases the information {about}  node locations could be shared   in space networks. If this information is eavesdropped, connections could be exposed to major security threats. These issues need to be addressed when utilizing mobility protocols.  There are some efforts to investigate related aspects; for example,
 authors in~\cite{atiquzzaman2011security}   suggested various schemes to withstand mobility protocols security threats. However,   security and integration issues   require more investigation. 

CGR, due to its accuracy and efficiency, has been gaining attention recently and is shown to be the most effective routing protocol for space networks. Nonetheless, CGR faces many challenges, making it an intriguing area for researchers.
Despite the contributions and  efforts made in the topic, by  e.g.~\cite{fraire2021routing} and the references therein including~\cite{wang2016scoping},   CGR resiliency is still a concern and subject to enhancement. In support of this claim, improvements over CGR  have been shown to be possible, e.g. in~\cite{el2017eaodr} where the earliest arrival optimal delivery ratio (\acrshort{eaodr}) routing algorithm has been proposed. Being able to account for the earliest available  and future contacts, EAODR seems to offer  a 12.9\% delay reduction over CGR. 
In the same regard,  addressing CGR scalability and computational complexity that may incur due to  network size increase are still research concerns.  As a matter of fact, DTN scalability in general is still an open research challenge. 
Despite the efforts made in developing DS internetworking protocols,   
 more advanced  protocols that  account for  high delays, intermittent losses, network topology dynamicity, and links and entities heterogeneity while still guaranteeing high capacity are needed.
Furthermore, DS internetworking protocol needs to account for future services such as multicast/broadcast and  ensure efficient  satellite-to-Earth transmission to meet the demand for expanding human presence in the Solar system. Achieving redundancy by deploying more relaying systems orbiting over  different planets and moons can contribute toward achieving such a goal.

\begin{figure*}[t]
\centering
\includegraphics[width=.9\linewidth]{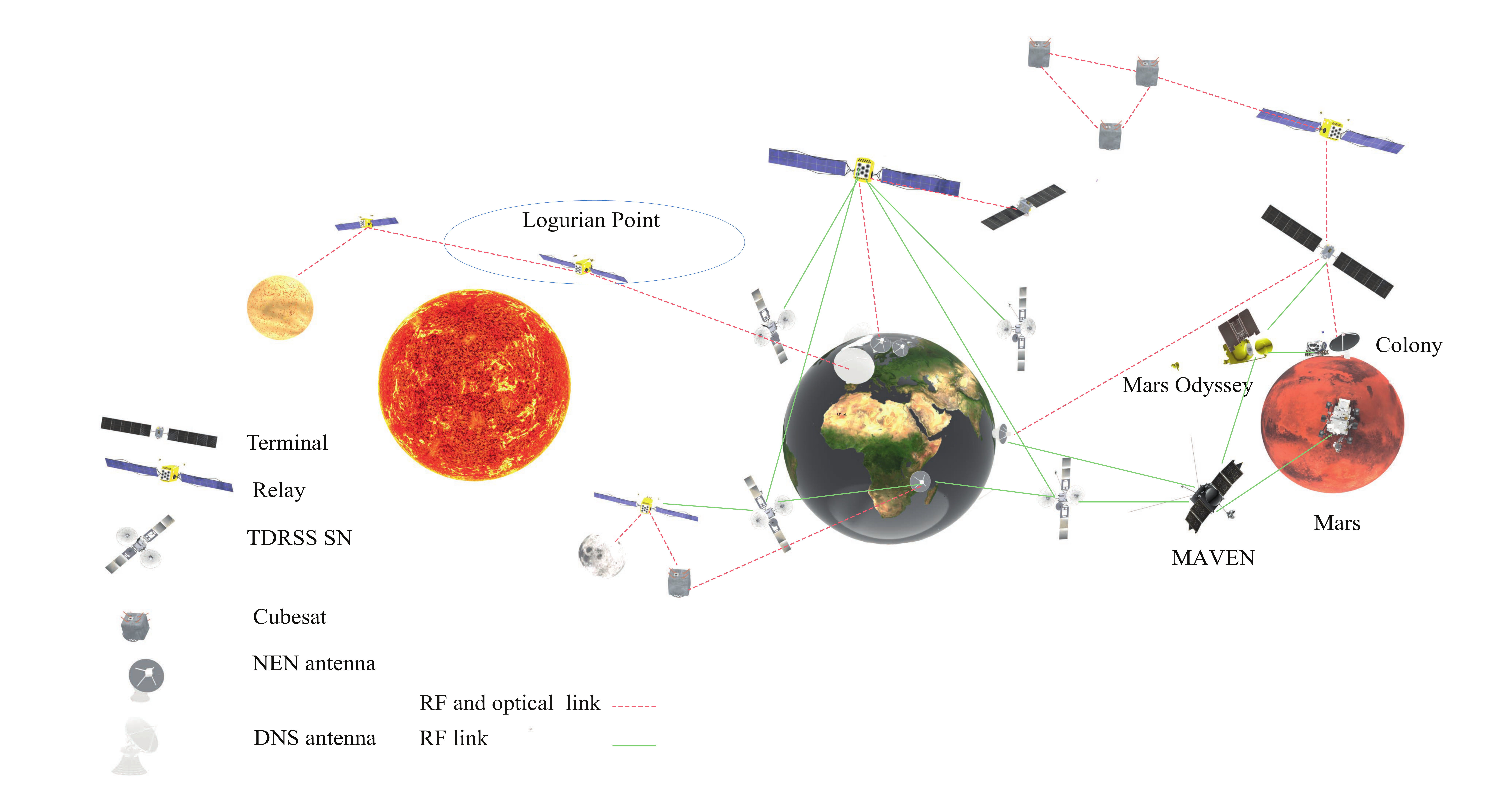}
\caption{Next generation IPN.}  
\label{ScaNnetwork}
\end{figure*}

\section{Advances OF DSOC}
\label{sec:dsAdvances}
Having discussed the key physical layer techniques and networking protocols along with their performance in the field of DSOC in Sections~\ref{sec:dsPMod}-\ref{sec:dscomd}, in this section we address a number of other emerging technologies and advancements that  promise to address DS communications challenges and meet future field goals. The section reviews some research and  implementation efforts done in the direction of the IPN, OAM, QC, and RF/FSO systems and identifies gaps to be explored. 

\subsection{Interplanetary Network}
NASA has three operational networks that  
 support missions: deep space network (DSN), near-Earth network (\acrshort{nen}), and space network (\acrshort{sn}). \acrshort{dsn} is the primary network that is responsible for communicating with majority of space missions.
While DSN consists of three stations distributed around Earth separated by 120{\degree} for full coverage, NEN consists of ground stations that provide communications and tracking services from low-Earth orbit to lunar orbit~\cite{roberts2014evolving}.
SN has a constellation of geosynchronous relays, which consists of 10 satellites called tracking and data relay satellite system (\acrshort{tdrss}),  along with two ground facilities.  SN will undergo over a redesign  of its architecture, as arranged by the  Space Network Ground Segment Sustainment (SSGS) project, to modernize the system to be capable of handling a few gigabits per second data rate~\cite{NASAgov2013SSGS}. 

 The goal of the \acrshort{ipn} is to provide connectivity all around the Solar system meeting the demand for high-data volumes including video streaming and high-quality images transmitted by the ever-increasing number of missions. The IPN will ultimately need to be accessible by the public~\cite{weber2006transforming}. Connecting  DSN, NEN, and SN networks is the first step in deploying the IPN.  
 
 Based on~\cite{reinhart2017enabling,alhilal2019sky,hurd2006exo},  we demonstrate in Fig.~\ref{ScaNnetwork}  the vision for the {IPN architecture.}  
Currently, through RF links, Mars and  Lunar entities have merely  access to the main Earth network. In the future, however, planets will have their own networks, and the IPN will become a network of networks. Each planet, as depicted in Fig.~\ref{ScaNnetwork}, will have a dedicated relay that offers a continuous connection to Earth through optical and RF-based links, referred to as  trunk links~\cite{reinhart2017enabling}.
Local services between planet elements are to be made possible through other, more available, optical and RF-based links, referred to as proximity links. The orbiting satellites around Mars, Mars Atmospheric and Volatile EvolutioN (\acrshort{maven}) and Mars Odyssey, shown in the figure, are all part of the current  Mars relay system. They are displayed to demonstrate Earth-Mars uplink and downlink communications. Terminal satellites  may act as a link to convert optical signals to RF and vice versa providing a significant advantage in overcoming atmospheric effects~\cite{hurd2006exo}. 
To avoid interplanetary link unavailability due to {obstruction} and Sun-Earth conjunction, spare interplanetary communications links through spacecrafts located at Lagrangian points are to be included~\cite{alhilal2019sky}. 

In addition {to implementing} FSO-based relays, which are  expected to offer 300 Mbps at  Mars distances for example~\cite{reinhart2017enabling},  Cubesats    are emerging,   as SWaP-reduced satellites 
 or  landers in some cases, to provide a 2 Gbps rate when thousands of them are implemented~\cite{velazco2020inter,saeed2020cubesat}.
The first two interplanetary Cubesats launched by  NASA in 2018, Mars Cube One (MarCO) probes, have successfully been deployed for navigation and monitoring~\cite{martin2019navigating}. 
The \acrshort{marco}s   mission, which is still operational, suggests more utilization of Cubesats for future missions~\cite{oudrhiri2020marco}.
Lately, to improve the IPN further, it has been suggested to use asteroids as interplanetary relays~\cite{kalita2018exploration}.
It is proposed to use Cubesat landers to form multiple antennas on the surface of an asteroid. Based on the numerical analysis provided in~\cite{kalita2018exploration}, an antenna of 400 m\textsuperscript{2} size provides up to 80 Mbps in Mars distances. Although Cubesat is a promising technology, networking issues resulting from long communications distances can become even more severe as the number of IPN nodes becomes large. For example, when Cubesats are implemented in large quantities, 
network protocols like CGR, and similarly EAODR, become inefficient~\cite{de2022dtn}. To mitigate that, however,  rate adaptation with a bandwidth fair sharing mechanism is suggested in~\cite{de2022dtn}, for example, to enhance the process of bundle transfer and management.
 
\subsection{Orbital Angular Momentum}
\label{OAM} 
To better utilize the  tremendous capacity offered by FSO, a great deal of attention is being given to exploring the potential of merging  FSO with OAM, which emerges  as a technology to  solve the high data rate  demands of future  terrestrial~\cite{trichili2019communicating,wang2012terabit,huang2014100,ren2016experimental,yousif2019performance,wang2015ultra} and DS communications~\cite{djordjevic2010ldpc,djordjevic2011deep,wang2014potential,zhang2022parallel,wang2022orbital}. 
OAM does not only have the potential to significantly enhance  spectral efficiency, but also to offer a great deal of  improvement 
over communications energy and security.
The light that {carries} OAM constructs a twisted beam with a helical phase front  making a singularity at the center of the beam resulting in a doughnut-like shape wavefront intensity profile.  
One of the OAM beam characteristics commonly exploited is the  topological charge or mode number, which is a positive or  negative integer that represents the number of phase twists per wavelength and the sign indicates the direction of the twist. Thanks to their orthogonality, OAM  modes with infinite distinguished phase fronts  provide  an additional  degree of freedom that results in significant capacity returns.
 {There are various OAM generation methods that are being used in the context of DS   including spatial light modulators (\acrshort{slm}s), spiral phase plate (\acrshort{spp}s)~\cite{zhang2022parallel},}
computer-generated holograms (\acrshort{cgh}s). 
These  schemes are also used for  detection. 
SLM, which could be followed by a charge-coupled device
(\acrshort{ccd}) camera when used as a detector~\cite{wang2014potential}, has shown to have several advantages 
over CGH and SPP schemes   including having a  higher-control accuracy and resolution~\cite{wang2014potential}.
SPPs, in addition,  have  {been} shown to have limitations when larger-mode numbers are considered. Accordingly, Zhang et al.~\cite{zhang2022parallel} suggested  {using the}  Dammann optical vortex grating (\acrshort{dovg}) method as it  can modulate more OAM modes than the SPP and generate independent collinear OAM channels~\cite{lei2015massive}.
Other methods including  holographic or interferometric  methods are recommended to be used for obtaining high-photon efficiency in DS~\cite{djordjevic2010ldpc,djordjevic2011deep}. 

{There is a handful of work done exploring  the performance of implementing OAM in DS. 
{From the capacity aspect,} 
 {Djordjevic et al.~\cite{djordjevic2010ldpc} showed that DS OAM link can achieve a rate of 100 Gbps, based on LDPC-coded multidimensional OAM  modulation.
  By employing   10 Gbps of data on 10 OAM  dimensions, that could be provided by using PPM for instance~\cite{djordjevic2011deep},  the aggregated rate of 100 Gbps has shown to be achievable. 
  Higher rates are accomplished by increasing the number of states used. 
 However,  scaling to higher states comes at the cost of increasing the crosstalk between the modes,  thereby lowering the performance of BER.
 %
 Looking at the OAM state-of-the-art technologies, data rates can  be in the range of hundreds of terabytes per second and even higher in some cases. Huang et al.~\cite{huang2014100}, for example, 
  showed that multiplexing quadrature amplitude modulation  (QAM) signals with OAM results in 100 Tbps for a  distance of 1 m. Wang et al.~\cite{wang2014n}, on the other hand, demonstrated that 54.139 Gbps orthogonal frequency division multiplexing
(OFDM) 8~\acrshort{qam} signals over 368 WDM pol-muxed 26 OAM
modes resulted in a capacity of 1.036 Pbps and spectral efficiency of 112.6 b/s/Hz. The authors in~\cite{wang2014n} have not related such a high-achievable rate to a  distance  though. 

Wang et al.~\cite{wang2014potential}  proposed  encoding OAM states of a single photon for DS communications.
They also discussed that the number of states {does} not affect  channel efficiency and hence, crosstalk is not of concern when  all communications entities are beyond the Earth's atmosphere. 
Higher-order states have been avoided in the analysis provided in~\cite{wang2014potential}  though, as they are more susceptible to diffraction. 
Furthermore, Wang et al.~\cite{wang2014potential} suggested that the deployment of relay nodes within OAM DS links to facilitate a reliable and secure transmission.
In~\cite{zhang2022parallel}, on the other hand,
a  weak beam intensity detection in  {the} presence of   high-background noise, measured in  counts per second (cps), has been realized using {a} single photon PC detector and  parallel coding. 
The proposed  technique, 
which is  referred to as  OAM-pair PPM coding,  
is suggested as an efficient method suitable for long distances including DS. 
Utilizing this scheme, each OAM-pair achieves $(\log_2(M)+1)$  bits per pulse resulting in    ${n}/{2}$ multiple of that total pulse density, when   $n$ even number of  OAM modes are  transmitted over  PPM signal with order $M$.  
According to their analysis, the authors demonstrated that $12\times$ increment in the capacity of PPM can be achieved. 
Regarding the BER performance, 
OAM-pair PPM 
achieves orders of magnitude improvement over the traditional PPM, as Fig.~\ref{OAMvsBG} demonstrates.


\begin{figure}[t!]
\centering
\includegraphics[width=0.85\linewidth]{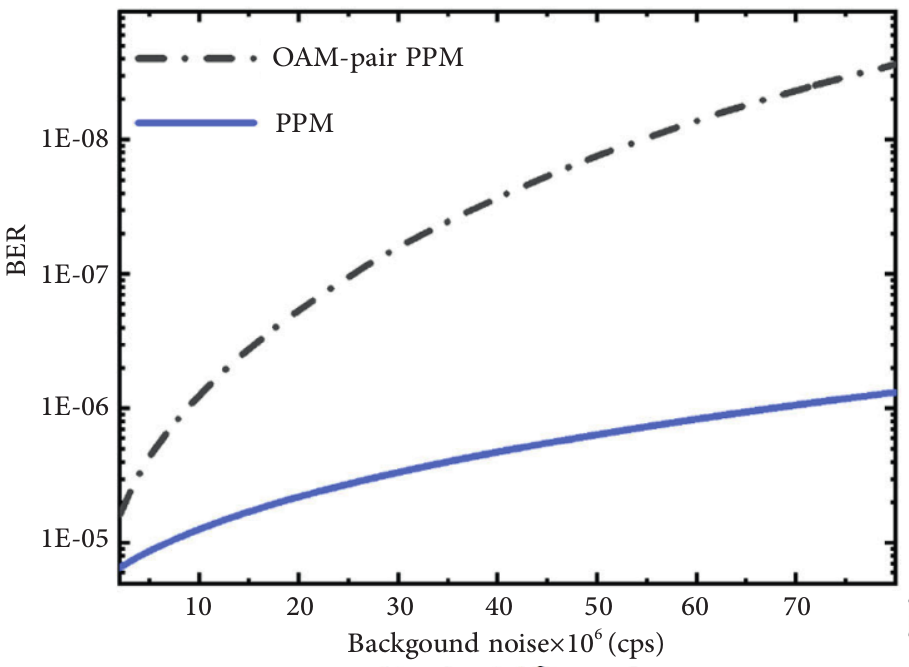}
\caption{Parallel coded OAM performance~\cite[Fig.~3]{zhang2022parallel}.}
\label{OAMvsBG}
\end{figure}

 {Many recent research and experimental work have, unprecedentedly, exhibited very significant results in the possibility of establishing high capacity. {Nonetheless, aside from~\cite{djordjevic2010ldpc,djordjevic2011deep,wang2014potential,zhang2022parallel}, to the best of our knowledge, no  more investigations have been dedicated in the context of DS.}
In addition to the general OAM implementation challenges~\cite{trichili2019communicating}, incorporating OAM in DS faces unique hurdles.  We below demonstrate these hurdles and discuss {the}  attempts to address  {a} number of them.}

\textbf{Atmospheric turbulence:} OAM modes can be subject to propagation effects, which can lead to crosstalk between  modes and affect the orthogonality between OAM states. 
Even though AT is not of concern in space-to-space links,  as stated previously, it is a significant challenge for  probe-to-Earth links. 
 When it comes to limited-distance communications, mainly up to several kilometers, one can find a variety of work done for compensating  {AT,} see e.g.~\cite{zhao2019orbital,dedo2020oam,hao2020high,li2020atmospheric,zhou2020high}}.
In~\cite{zhao2019orbital,dedo2020oam}}, for example, authors have shown, through the use of deep learning techniques, {that} 100\% OAM mode recognition is possible in medium to weak turbulence levels, while between 50\%~\cite{zhao2019orbital} and 80\%~\cite{hao2020high} recognition can be achieved in the case of strong turbulence. 
Zhao et al.~\cite{zhao2019orbital} used  {an} image classification based diffractive neural network to detect 10 OAM modes under AT. 
They indicated that at a significant level of turbulence, detection accuracy drops very low due to the turbulence-induced spiral spreading that causes crosstalk between the OAM modes posing detection limitations.
Furthermore, Hao et al.~\cite{hao2020high},  used a conventional neural network to detect 16 OAM modes,
under various turbulence levels based on image processing while taking into account the complexity and computation time to ensure  practicality. 
It is relevant to consider AT for DS links due to the use of ground entities in many scenarios.
However,  the success of  deep learning and image classification in  addressing AT in the context of DS has never been explored,  {to} the best of our knowledge. 
Other  methods like  correction codes could also be implemented to mitigate AT. 
In the context of DS, Djordjevic et al.~\cite{djordjevic2010ldpc} proposed combining the LDPC code with OAM-based modulation schemes to overcome AT effects.
Considering the DS links AT mitigation methods presented in~\cite{djordjevic2010ldpc}, 
 which are also recommended for near-Earth communications, the authors of~\cite{djordjevic2011deep} shown the possibility of the  system  to {withstand strong turbulence} and achieve a spectral efficiency of $n^{2}/\log_2(n)$ times that of LDPC-PPM modulation for $n$  number of  modes. Noting that~\cite{djordjevic2010ldpc,djordjevic2011deep} are the only efforts attempted in regard to taking AT into consideration for DS.

\textbf{Spatial spread:} {OAM properties can be utilized in different types of beams. An Example of such beams is Laguerre-Gaussian (\acrshort{lg})  which have the self-healing property, propagates {over long} distances~\cite{mendoza2015laguerre}, and can be generated easily. Owning to the prementioned attributes, LG is preferred  over other types~\cite{wang2014potential,djordjevic2011deep}.} 
   LG beam's intensity profile exhibits a doughnut-like shape that gets bigger, i.e., spreads or diverges more spatially as they travel,  making their detection over DS distances challenging.
 In Fig.~\ref{OAMmodes1},  we show the LG beam intensity profile and spatial distribution  for modes $L=0,\Ou1$ ,$\Ou2$,   transmitted with  {a} 1 m diameter antenna, at  {a}  distance of 401$\times$10\textsuperscript{6} km, which is chosen based on the longest distance to Mars.
 We can notice from the figure that  OAM beams preserve their shape even after  traveling  very long distances and that beams spread more as the mode number increases.  The spread and shape of different OAM modes are of crucial importance when it comes to locating DS communications nodes so it is guaranteed that the singularity, in the middle of a beam, is avoided. 
 
 The good news, though, that could make reception possibly reasonable at ultra-long distances, despite the huge spatial spread,  is the fact that OAM  modes {have} been experimentally proven  to  be detected even if they are  partially received~\cite{xie2013analysis,elhelaly2018reduced,deng2019orbital,zheng2015orbital}. 
Modes detection through receiving a very small portion of the whole circumference of {an} electromagnetic wave carrying OAM has been reported by~\cite{shi2023demodulation}.
Surprisingly, partially receiving a beam can lead to a reduction in crosstalk effects, surpassing the performance of full area receivers in some cases. 
Furthermore, according to~\cite{zheng2015orbital}, a limited size full area receiver that does not capture the maximum of the OAM mode intensity profile is  outperformed by   {a} partial receiver  positioned at the maximum intensity of the beam. 
As indicated in~\cite{elhelaly2018reduced},  for two overlapping beams, two  detectors receiving   three  {quarters} of the  sectional area at an SNR $30$\,dB   achieve  $12$\,b/s/Hz higher-spectral efficiency    than a complete area receiver.
 Deng et al.~\cite{deng2019orbital}, in addition,  proposed an approach that resulted in a 46.2\% reduction in crosstalk intensity, achieving  as low as 7.71\%  with   four  non-adjacent OAM states and no more than 22\% for  two adjacent modes, indicating that closer mode numbers induce more crosstalk.
  Despite the OAM partial reception promises shown in limited communications distances,  however, as beams {are} largely spread in DS, the minimum portion of a beam that {is} needed for a mode to be recognized could be relatively large. To the best of our knowledge, the partial signal detection of OAM beams at DS distances, has never  yet been  investigated.  Analyzing its feasibility using   conventional methods as well as artificial intelligence based schemes would be of interest.}}
  
\textbf{Lateral displacement error:} 
Keeping an LG beam axis and its receiving entity fully aligned is difficult.  For long-distance transmissions, this requirement is even more challenging. 
An alignment error does not only render optical power loss, but also,  more crucially, cause expansion of the spiral spectrum resulting in {a} crosstalk effect and mode detection difficulty~\cite{willner2016design}. Even though it has been shown that the lateral offset can improve resiliency against eavesdropping~\cite{gibson2004OAMsecurity}, and can be mitigated with the use of lateral correction methods~\cite{zhang2021extraction}, {such an analysis is lacking in DS communications.}

Despite the efforts {made}, a significant contribution  still needed to handle DS challenges considering the accomplishments {made}  in limited distances  moving forward to facilitating OAM implementation in DS.

\begin{figure}[t]
\centering
\includegraphics[width=1.0\linewidth]{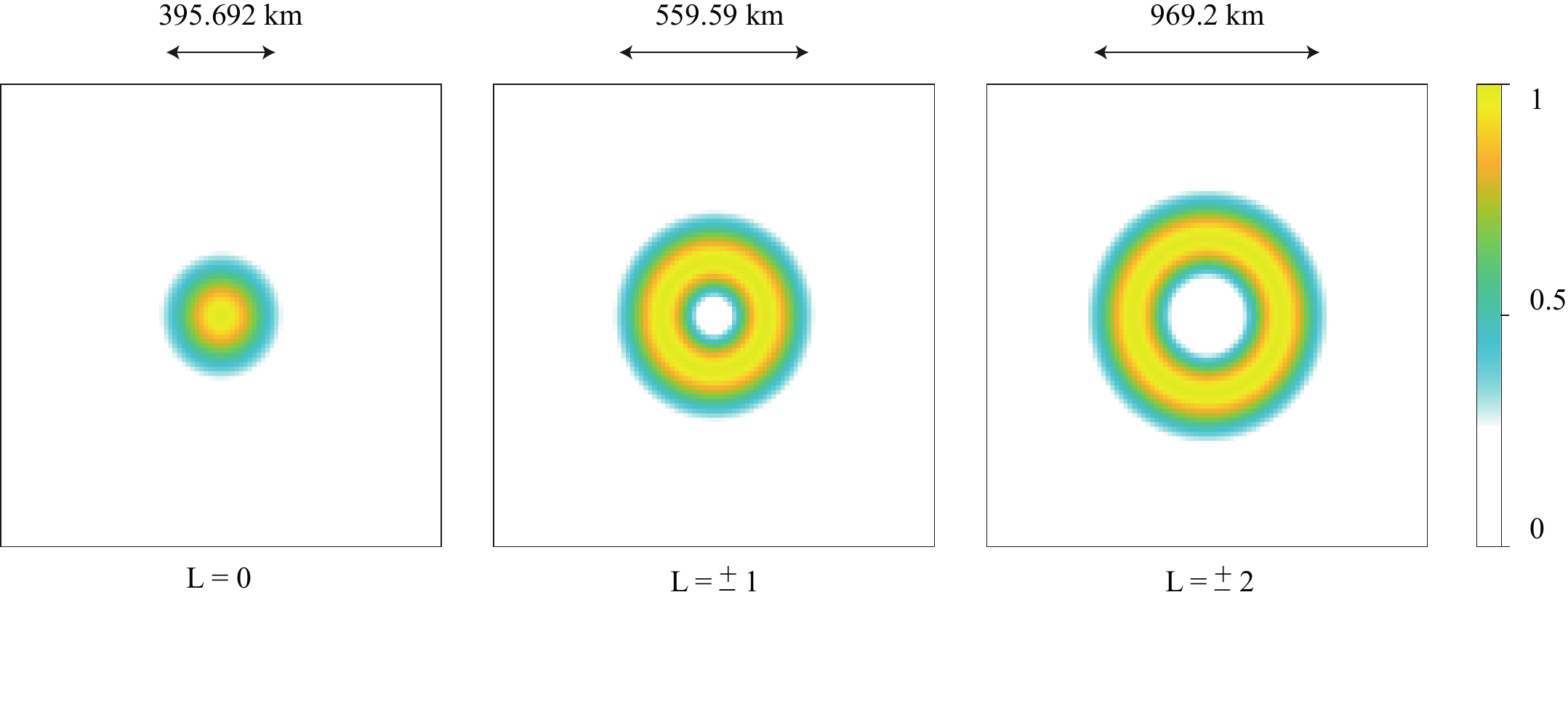}
\caption{OAM beam dimensions and intensity at the farthest Earth-to-Mars distance.}
\label{OAMmodes1}
\end{figure}

\subsection{Quantum Communications} 
\label{Quanta}
 The quantum technology is expected to break through the limits of classical  technologies in the aspects of  security, computational capabilities, as well as information transmissions. Quantum technology is marked to provide an absolute secure channel. In addition, theoretical and experimental results have shown that QC can increase the capacity of classical channels with methods such as, superdense coding~\cite{bennett1992communication,kwiat2016entanglement} and superadditivity~\cite{hastings2009superadditivity,kwiat2016entanglement}.
 Quantum properties have not only  gained  attention for being used in terrestrial  communications~\cite{zhang2019quantum}, but also in space~\cite{zhang2019quantum} and DS~\cite{antsos2020jpl}.
 Interestingly, based on current technology, involving entities from   space for achieving terrestrial QC of distances beyond a hundred kilometers is not optional as it provides the means to avoid the hurdles caused by the  Earth's atmosphere. 
 Furthermore, unlike terrestrial mediums  where quantum  states are susceptible to  uncorrected changes~\cite{mohageg2111deep}, because there are negligible channel effects on signals traversing space, it is expected that quantum properties such as entangled photons polarization and other degrees of freedom  would persist in DS. 
 The suitability of the  DS environment for  QC is highlighted in the   work~\cite{kwiat2016entanglement,chapman2018progress,rengaswamy2021belief,delaney2022demonstration}, where the possibilities of increasing  communications rate over DS  by the involvement of  quantum links are discussed.
  
  Among the ongoing initiatives for establishing DS QC is the Deep Space Quantum Link (\acrshort{dsql}) project~\cite{antsos2020jpl},  which is intended to conduct several unprecedented quantum experiments at long distances including observing  gravitational effects on quantum optical systems, and performing the Bell test, teleportation, and entanglement swapping.
It is anticipated that within the next decade, a full quantum teleportation from Earth to {the} Lunar Gateway, which is a space station that would be orbiting the Moon, is expected to  be achieved~\cite{hutopical}. 
To realize such  objectives, various technological challenges specifically relating to   quantum properties that are not yet practical are required to be tackled. 
The development of DSQL will consequently pave the way for space quantum communications and the realization of a global quantum network~\cite{mazzarella2021goals}.
Kwiat et al.~\cite{kwiat2016entanglement} attempted to make QC possible in DS, taking advantage of the less lossy channels with superdense coding, which could enhance the classical communications capacity by a factor of two without the need for preparing entanglement.
The realization of {the} superdense teleportation protocol (\acrshort{sdt}), proposed in~\cite{kwiat2016entanglement} as a candidate for DS, that is implementable in high fidelity is in progress~\cite{chapman2018progress}. 
Rengaswamy et al.~\cite{rengaswamy2021belief}  demonstrated that a quantum-optimal joint detection receiver, which is based on the belief propagation with quantum messages (\acrshort{bpqm}) algorithm~\cite{renes2017belief}, through the superadditivity concept can enhance classical communications. 
Such a receiver  detects complete codewords for BPSK {pure-loss} channel getting  to the possibility of attaining {the} Holevo limit, hence  becoming suitable for DSOC. 
In the same direction,~\cite{delaney2022demonstration} showed that joint detection receivers outperform  one at a time codeword BPSK demodulators  reaching five times {the} classical decoders  rate.

Furthermore, on a fundamental aspect, there are ongoing studies on the use of {the} Aharonov-Bohm effect in DS, a quantum  phenomenon in which the electromagnetic potential affects particles charge, which promises  the possibility of superluminal communications and  interplanetary QC. 
If the application of the effect is realized, {the resulting} communications capabilities {are} anticipated to surpass conventional optical communications~\cite{oleinik2003information,bushnell2023survey}. 
Through performing quantum teleportation with X-rays serving as the quantum information carrier, beyond interplanetary links,  interstellar communications {are} possible according to~\cite{filipova2023interstellar,berera2022viability}.
The aforementioned evidence of the possibility of upgrading  DS communications performance {beyond} the limits of classical methods could render a new set of applications and techniques that mitigate  significant DS challenges.

\subsection{Radio Frequency/Free Space Optical Systems}
\label{sec:hybrid}
 {Even though the FSO potentials in DS are tremendous, full implementation of the  technology is  deferred.   RF technology is still dominating space communications facilities, as  discussed in Section~\ref{sec:dscome}. 
 Some efforts in the literature support the idea of combining RF and FSO in DS  as a starting point. 

The two schemes for integrating these technologies are hybrid RF/FSO and dual-hop RF/FSO. 
While a hybrid RF/FSO system combines the RF and  FSO technologies in parallel, dual-hop RF/FSO {implements} them in series.  
The ability of hybrid RF/FSO to switch between RF and FSO leads to overall DS system performance improvement~\cite{trichili2021retrofitting} as it does in terrestrial~\cite{moradi2010availability}. 
 {The} RF/FSO dual-hop method, on the other hand, increases coverage, enhances DS communications reliability, and lowers the overall system cost.
In~\cite{xu2020dual}, for example,  the authors demonstrated the benefits of using dual-hub RF/FSO in lowering the probability of link outages in the presence of Solar scintillation. Similarly, authors of~\cite{xu2021mixed} analyzed the performance of RF/FSO systems and showed that they outperform both  FSO and RF links in weak to moderate coronal turbulence. They also showed that  BPSK modulation outperforms some other commonly used schemes in such an environment.~\cite{djordjevic2017oam} has also shown that combining RF/FSO with OAM increases spectral efficiency. To boost capacity and reduce turbulence effects  in DS, furthermore,~\cite{yousif2019atmospheric} suggested combining OAM multiplexing in a \acrshort{mimo} system,  {which operates with modified PPM modulation with spatial mode multiplexing in a hybrid RF/FSO system.}

{There are, besides, dedicated efforts to make RF/FSO operational for  DS. The Integrated Radio and Optical Communications (\acrshort{iroc}) project~\cite{raible2014physical}, for example,   features a hybrid RF/FSO aperture antenna,  called 
Teletanna, that  reduces SWaP burden and utilizes a beacon-less and high data rate system. iROC is anticipated to be operational around Mars~\cite{melis2022iroc} 
with SWaP comparable to Mars Reconnaissance Orbiter (\acrshort{mro}) and data rate of 267 Mbps, which is orders of magnitude higher than current  rates achievable by the RF technology. Furthermore, if the same technology is used for lunar distances a 10 Gbps data rate can be achieved.} 

\subsection{Discussions and Outlook}
 The term IPN is now used to categorize networks that connect several spacecrafts placed in different regions of space. For example, a network consisting of several  Mars and Earth satellites is referred to as a third IPN~\cite{hurley2003current}. However, the envisioned IPN is a network that connects all planets elements analog to the Internet on Earth reliably. The IPN will need to handle the growth of the multimedia data volume~\cite{wang2013qos} and support the human presence in DS. 
 
 OAM is getting a lot of attention outside the scope of DS as  a strong candidate for the upcoming big data era. 
  OAM has a remarkable performance in fiber optics as it can offer a large capacity for long-distance transmission~\cite{hassan2022novel}; the number of modes the fiber can handle, for example, can reach 394~\cite{fu2022photonic}. When it comes to wireless communications, nonetheless,  OAM  performance can be significantly impacted by signal transmission distance,  atmospheric turbulence, and the ability to detect partially received beams. Resulting possibly in  trading off  the   increase in OAM dimensionality and hence data rate with achieving  quality performance. 
 More investigation and maturity for the technology will be needed until it becomes applicable to DS. 

Among the other advantages~\acrshort{qc} offers, the improvements they promise in terms of capacity, making them attractive to be implemented in DS to overcome the ultra-long  distance losses and meet the demand for the high rate. It will be even more interesting to integrate, for channel multiplexing, OAM with QC. A similar idea has been implemented  for a fiber optic~\cite{zahidy2022photonic}  and   space~\cite{wang2019detecting} 
 links for capacity boosting.  
Scalable QC techniques are significantly challenging to realize, mainly due to the fact that quantum repeaters, generally speaking, are challenging to deploy~\cite{chou2007functional,yuan2008experimental,kalb2017entanglement,hensen2015loophole}, as they require high accuracy  {in} quantum measurements and operation  and  memory efficiency.   
Despite the attempts that promise the practicality of quantum repeaters~\cite{bhaskar2020experimental} and memory enhancement, 
interestingly, however, studies have  shown  deploying \acrshort{qc} over Earth-to-space   links reaching a distance of 1200  km  {has} been possible  without the need of repeaters~\cite{villoresi2008experimental,yin2017satellite},  as they suffer less than their terrestrial counterparts from turbulence. 
As the turbulence effects are even less between space-to-space links, it would be of interest  to expand such experimental studies to  DS given the fact that utilization of QC over space links is preferred.

RF/FSO systems have shown to be an effective method to offer  reliability enhancement and communications improvement, for example, in terms of  BER and outage probability~\cite{bag2018performance}, spectral efficiency~\cite{djordjevic2017oam}, and link outage probability~\cite{xu2020dual}.
The gain of the technology merge seemed to be encouraging  the space industry in the  start of implementing  FSO beyond the scope of research and development, e.g.  in iROC and  Mars Perseverance rover local communications~\cite{ProvencesMars2020}, 
to meet the future demand for DS connectivity and capacity enhancement, especially when considering the RF technology  maturity and  RF-to-FSO transition cost. 

\section{Concluding Remarks}\label{sec:conc}

\subsection{Lessons Learned}
Existing RF systems are unable to meet future DS communications demands as opposed to {their} FSO counterpart. 
Rendering to the FSO wavelength characteristics, {which provide an abundance} of bandwidth, low beam divergence and high directivity  {can be achieved}. 
Consequently, FSO  promises significant improvements in system performance including lower SWaP requirements and substantially higher-capacity returns than RF. 
However, implementing FSO  still faces  major impediments.
Starting off with the physical layer, there are three primary challenges,  namely handling the strong power dissipation, pointing loss, and atmospheric effects. 
While the first can be tolerated by obtaining high-gain transmitters and receivers, pointing loss can be tackled by aid tracking beacons.
The space-to-space links can be promising not only to avoid AT effects, but also  help to realize {OAM},  global  quantum communications, and  IPN. 

Modulation, coding, and detection schemes play 
important roles  {in} mitigating DS environment impairments. 
As concluded, PPM scheme is highly recommended {and commonly used} among the DSOC research community. 
 PPM has a high peak-to-average power ratio which leads to capacity improvement. 
Nevertheless,   PPM  has performance limitations  when higher-data rates are demanded. In addition, increasing PPM order results in higher-system complexity and cost. 
Furthermore,  the capacity limit analysis suggests to investigate other modulation schemes than PPM.    When it comes to coding,  methods  that are  well studied in DS   are limited and mostly  restricted to PPM.
The detection of a DSOC system significantly  affects the capacity limit and modulation performance outcomes. The more sensitive the detector the better it is. However, sensitivity could inherently be limited by the choice of modulation, coding, and quantum noise.
Although the trend recommends direct detection techniques, coherent detection methods are gaining more attention currently as they achieve  {higher}-information rates.

Moving  to the networking aspect,  DTN is the most recommended architecture in DS mainly defined by the bundle layer and its related addressing mechanism. 
However, more attention is needed to enhance the DS protocols to achieve Solar system exploration  goals and  ultimately universify the network, for which, IPN is certainly one of the solutions.
On the other hand, we believe that Cubesats technology will be a major contributor {to} accomplishing the future IPN. 
Moreover, OAM is a promising technology for achieving data rates around  hundreds of gigabytes. 
Unlike other technologies, there is a gap in the literature in terms of applying OAM in DS. 
Moreover, partial detection, pointing and displacement, and AT  challenges are yet  to be overcome.
Acknowledging that the objectives to realize a global quantum network could be achieved by utilizing space links should motivate more investment in DS QC, which in turn could provide a new perspective on the technology  employment in general.
Merging RF and FSO in DS communications, which has been partially applied in DS, could be the starting point for applying FSO.  RF/FSO systems are shown to perform better than  FSO  in a number of cases.

\subsection{Summary}
The success of space exploration missions relies heavily on the reliability and efficiency of DS communications. 
That is achieved through different technologies and protocols, which can overcome the unique challenges of DS. 
In this paper, we provided an overview of  different technologies and discussed related challenges and possible solutions.  
We presented a detailed discussion on the  DSOC physical layer characteristics, physical performance analysis, protocols, and  field advances. 
We  {addressed} the future requirements that need to be fulfilled to cope with the changes in space missions, including compromising between  {mature} existing technologies and recent advances. 
The paper should provide a basis for researchers interested in learning about and contributing {to the} underlying technologies.

\section*{Acknowledgement}
The authors would like to thank Abderrahmen Trichili, 
and Bassem Khalfi for their valuable {input} and feedback for the paper.

\printglossary[style=modsuper, type=\acronymtype]

\bibliographystyle{IEEEtran}
\bibliography{2References}

\end{document}